\documentclass{elsart}
\usepackage {graphicx}
\usepackage{amssymb}
\input epsf

\newcommand{\D}{{\rm d}}

\newcommand{\uu}{\mbox{\boldmath$u$}}

\newcommand{\vv}{\mbox{\boldmath$v$}}

\newcommand{\yy}{\mbox{\boldmath$y$}}
\newcommand{\xx}{\mbox{\boldmath$x$}}

\begin{document}

\begin{frontmatter}
\title{Quasi-Equilibrium Closure Hierarchies for the Boltzmann Equation}

\author{Alexander N. Gorban\corauthref{cor1}}
\ead{ag153@le.ac.uk} \corauth[cor1]{Corresponding author:
Department of Mathematics, University of Leicester, LE1 7RH
Leicester,  UK}
\address{Centre for Mathematical Modelling,
University of Leicester, UK \\ and Institute of Computational
Modeling SB RAS, Krasnoyarsk, Russia}

\author{Iliya V. Karlin}
\ead{karlin@lav.mavt.ethz.ch}
\address{ETH Z\"{u}rich, Institute of Energy Technology, CH-8092
Z\"{u}rich, Switzerland }

\date{}
\maketitle

\begin{abstract}
In this paper, explicit method  of constructing  approximations
(the Triangle Entropy Me\-thod) is developed for nonequilibrium
problems.  This method enables one to treat any complicated
nonlinear functionals that fit best the physics  of a problem
(such  as, for  example, rates of processes) as new independent
variables.

The work of the method was demonstrated on the Boltz\-mann's--type
kinetics. New   macroscopic variables are introduced  (moments  of
the Boltzmann  collision integral, or scattering rates). They are
treated  as  independent variables rather than as infinite moment
series. This  approach gives the complete  account  of  rates  of
scattering  processes. Transport equations for scattering rates
are obtained (the  second hydrodynamic chain), similar to the
usual moment chain (the  first hydrodynamic chain). Various
examples of the closure of the first, of the second, and of the
mixed hydrodynamic chains  are considered for the hard spheres
model. It is  shown, in particular, that the complete account  of
scattering processes  leads to a renormalization of transport
coefficients.

The method gives the explicit solution for the closure problem,
provides thermodynamic properties of reduced models, and can be
applied to any kinetic equation with a thermodynamic Lyapunov
function.

\end{abstract}


\end{frontmatter}

\section*{Introduction}

In this paper, explicit method  of constructing  approximations
(the Triangle Entropy Me\-thod \cite{MBCh,MBChLANL}) is developed
for nonequilibrium problems  of Boltz\-mann's--type kinetics, i.e.
when the standard moment variables become insufficient. This
method enables one to  treat any complicated nonlinear functionals
that fit best the  physics  of a problem (such  as, for  example,
rates of processes) as new independent variables.

The method  is  applied  to  the  problem  of  derivation  of
hydrodynamics  from  the  Boltzmann equation.   New   macroscopic
variables are introduced  (moments  of  the  Boltzmann  collision
integral, or scattering rates). They are treated  as  independent
variables rather than as infinite moment series.  This  approach
gives the complete  account  of  rates  of  scattering  processes.
Transport equations for scattering rates are obtained (the  second
hydrodynamic chain), similar to the usual moment chain (the  first
hydrodynamic chain).

Using  the  triangle  entropy  method,  three different types of the
macroscopic  description  are  considered. The first type  involves
only moments of distribution  functions, and results coincide with
those of the Grad method in the Maximum Entropy version. The second
type of description involves only scattering rates.  Finally, the
third type involves both the moments and the scattering rates  (the
mixed description).

The second and the mixed hydrodynamics are sensitive to the choice
of the collision model.  The  second hydrodynamics is equivalent to
the first hydrodynamics only for Maxwell  molecules, and the mixed
hydrodynamics exists for  all  types of collision models excluding
Maxwell molecules.

Various examples of the closure of the first, of the second, and of
the mixed hydrodynamic chains  are considered for the hard spheres
model. It is  shown, in particular, that the complete account  of
scattering processes  leads to a renormalization of transport
coefficients.

The method gives the explicit solution for the closure problem,
provides thermodynamic properties of reduced models, and can be
applied to any kinetic equation with a thermodynamic Lyapunov
function, for example, to the Fokker--Planck Equation.

\begin{figure}
\begin{centering}
a)\includegraphics[width=60mm, height=30mm]{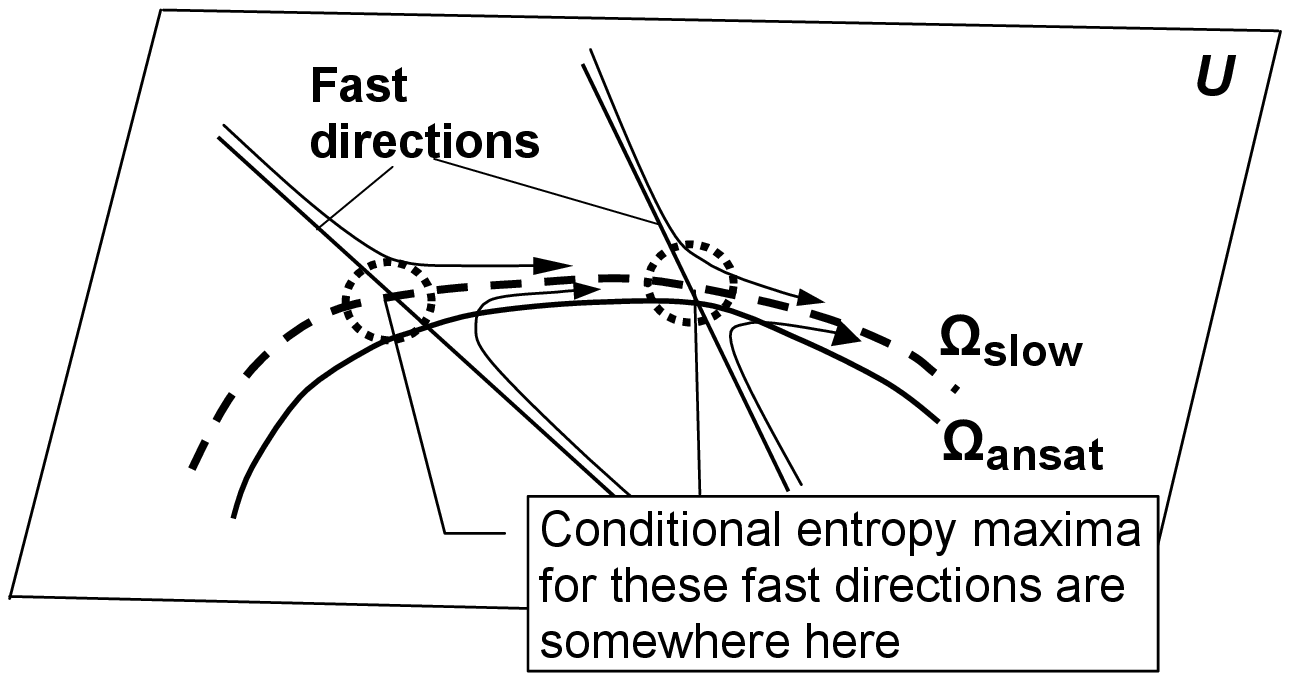}
b)\includegraphics[width=60mm, height=30mm]{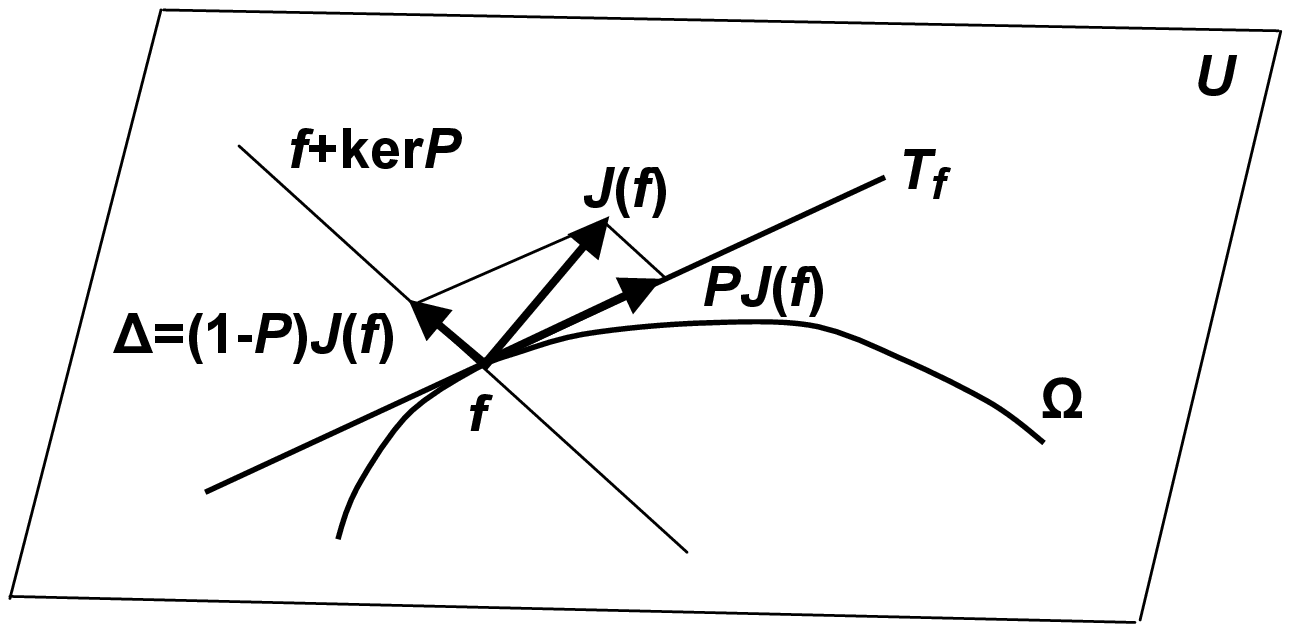}
\caption {\label{fig1QEHier} a) Fast--slow decomposition. Bold
dashed line -- slow invariant manifold; bold line -- approximate
invariant manifold; several trajectories and correspondent
directions of fast motion are presented schematically.  b) The
geometrical structures of model reduction: $U$ is the phase space,
$J(f)$ is the vector field of the system under consideration: $\D
f / \D t = J(f)$, $\Omega$ is an ansatz manifold, $T_f$ is the
tangent space to the manifold $\Omega$ at the point $f$, $PJ(f)$
is the projection of the vector $J(f)$ onto tangent space $T_f$,
$\Delta = (1-P)J(f)$ is the defect of invariance, the affine
subspace $f+ \ker P$ is the plain of fast motions,  and $\Delta
\in \ker P$.}
\end{centering}
\end{figure}

Reduction of description for dissipative kinetics assumes
(explicitly or implicitly) the following picture
(Fig.~\ref{fig1QEHier}a): There exists a manifold of slow motions
$\Omega_{\rm slow}$ in the space of distributions. From the
initial conditions the system goes quickly in a small neighborhood
of the manifold, and after that moves slowly along it. The
manifold of slow motion (slow manifold, for short) must be
positively invariant: if a motion starts on the manifold at $t_0$,
then it stays on the manifold at $t>t_0$.  In some neighborhood of
the slow manifold the directions of fast motion could be defined.
Of course, we always deal not with the invariant slow manifold,
but with some approximate (ansatz) slow manifold $\Omega$.

There are three basic problems in the model reduction:
\begin{enumerate}
\item{How to {\bf construct} an approximate slow manifold;}
\item{How to {\bf project} the initial equation onto the constructed approximate slow manifold, i.e. how to split motions
into fast and slow;}
\item{How to {\bf improve} the constructed manifold and the projector in order to make the manifold more invariant and
the motion along it slower.}
\end{enumerate}

The first problem is often named ``the closure problem", and its
solution is the closure assumption; the second problem is ``the
projection problem". Sometimes these problems are discussed and
solved simultaneously (for example, for the quasiequilibrium, or,
which is the same, for MaxEnt closure assumptions
\cite{Janes1,Zubarev,KoRoz,Ko,Roz,Kark}). Sometimes solution of
the projection problem after construction of ansatz takes a long
time. The known case of such a problem gives us the
Tamm--Mott-Smith approximation in the theory of shock waves (see,
for example, \cite{GK1}). However if one has constructed the
closure assumption which is at the same time the {\it invariant
manifold} \cite{GK1,GKTTSP94,Lam}, then the projection problem
disappears, because the vector field is always tangent to the
invariant manifold. In this paper, we would like to add several
new tools to the collection of methods for solution the closure
problem. The second problem was discussed in Ref. \cite{UNIMOLD}.
We do not discuss here the third main problem of model reduction:
How to improve the constructed manifold and the projector in order
to make the manifold more invariant and the motion along it more
slow. This discussion can be found in various works
\cite{GK1,GKTTSP94,Lam,IneManCFTe88,JonTiti}, and the broad review
of the methods for invariant manifolds construction was performed
in Refs. \cite{CMIM,GKSpri}.

Our standard example in this paper is the Boltzmann equation, but
most of the methods can be applied to an almost arbitrary kinetic
equation with a convex thermodynamic lyapunov function. Let us
discuss the initial kinetic equation as an abstract ordinary
differential equation\footnote{Many of partial differential
kinetic equations or integro-differential kinetic equations with
suitable boundary conditions (or conditions at infinity) can be
discussed as abstract ordinary differential equation in
appropriate space of functions. The corresponding {\it semigroup}
of shifts in time can be considered too. Sometimes, when an
essential theorem of existence and uniqueness of solution  is not
proven, it is possible to discuss a corresponding shift in time
with the support of physical sense: the shift in time for physical
system should exist. Benefits from the latter approach are obvious
as well as its risk.},
\begin{equation} \label{sys}
{\D f \over \D t}= J(f),
\end{equation}
where $f=f(q)$ is the distribution function, $q$ is the point in
particle phase space (for the Boltzmann equation), or in
configuration space (for the Fokker-Planck equation). This equation
is defined in some domain $U$ of a vector space of admissible
distributions $E$.

The dissipation properties of the system (\ref{sys}) are described
by specifying the \textit{entropy} $S$, the distinguished
\textit{Lyapunov function} which monotonically increases along
solutions of equation (\ref{sys}). In a certain sense, this
Lyapunov function is more fundamental than the system (\ref{sys})
itself. That is, usually, the entropy is known much better than
the right hand side of equation (\ref{sys}).

We assume that a concave functional $S$ is defined in $U$, such
that it takes maximum in an inner point $f^*\in U$. This point is
termed the equilibrium.

For any dissipative system (\ref{sys}) under consideration in $U$,
the derivative of $S$ due to equation (\ref{sys}) must be
nonnegative,
\begin{equation}\label{sign}
  \frac{ \,\D S}{ \,\D t}\bigg|_f=(D_fS)(J(f))\geq0,
\end{equation}
where $D_fS$ is the linear functional, the differential of the
entropy.

\begin{figure}
\begin{centering}
a)\includegraphics[width=60mm, height=40mm]{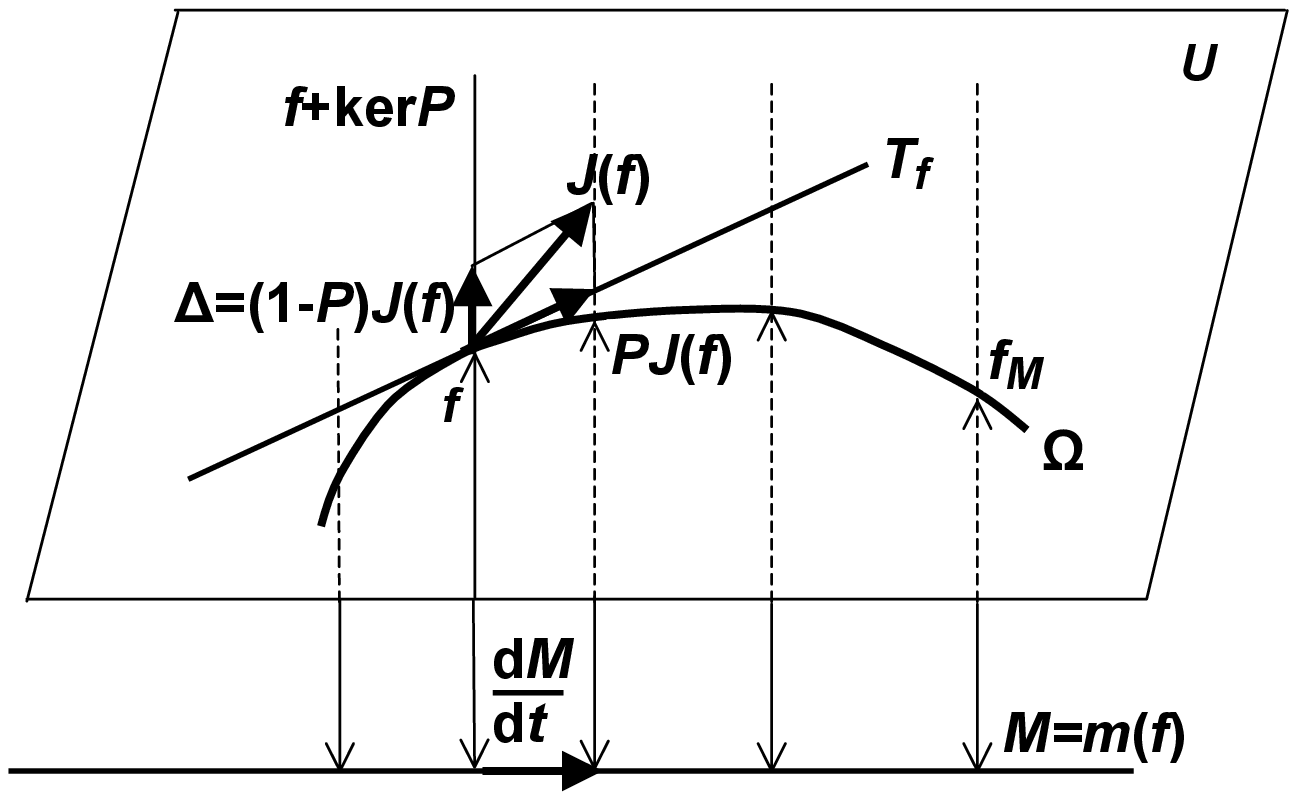}
b)\includegraphics[width=60mm, height=40mm]{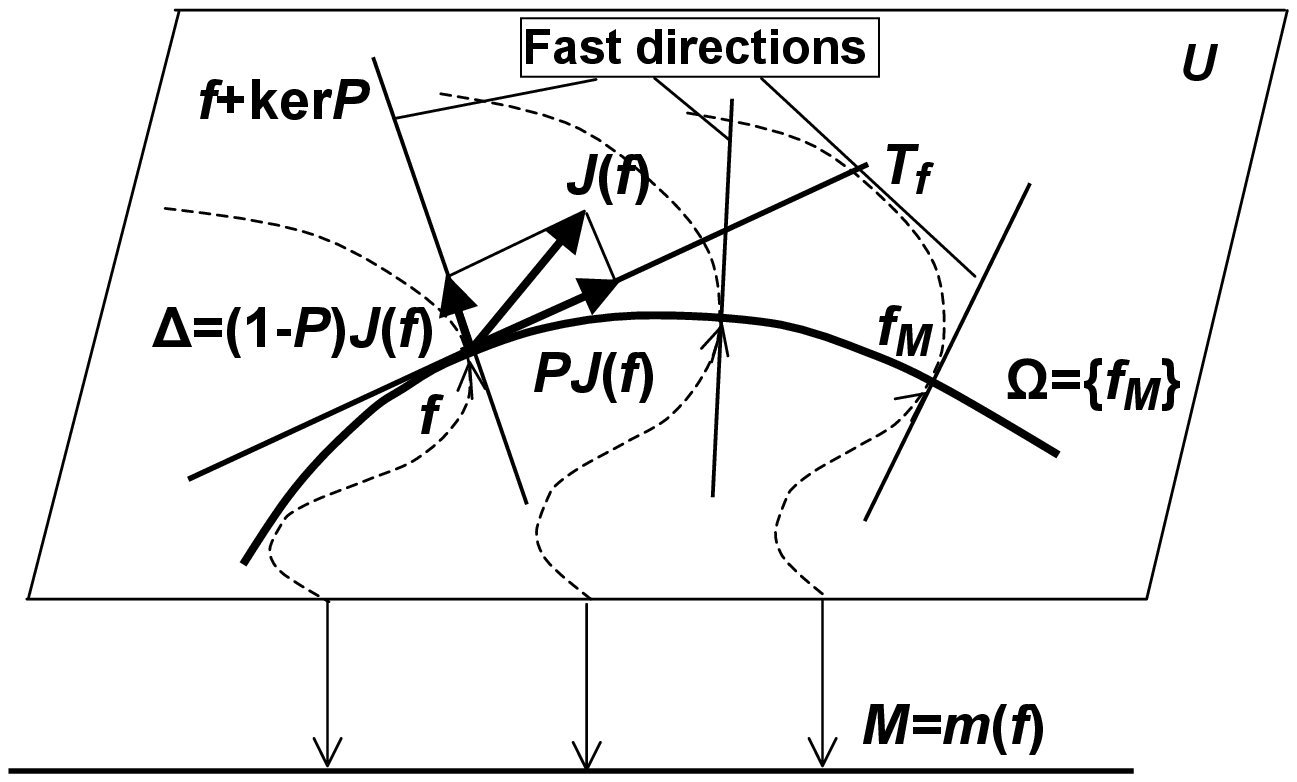} \\
c)\includegraphics[width=60mm, height=40mm]{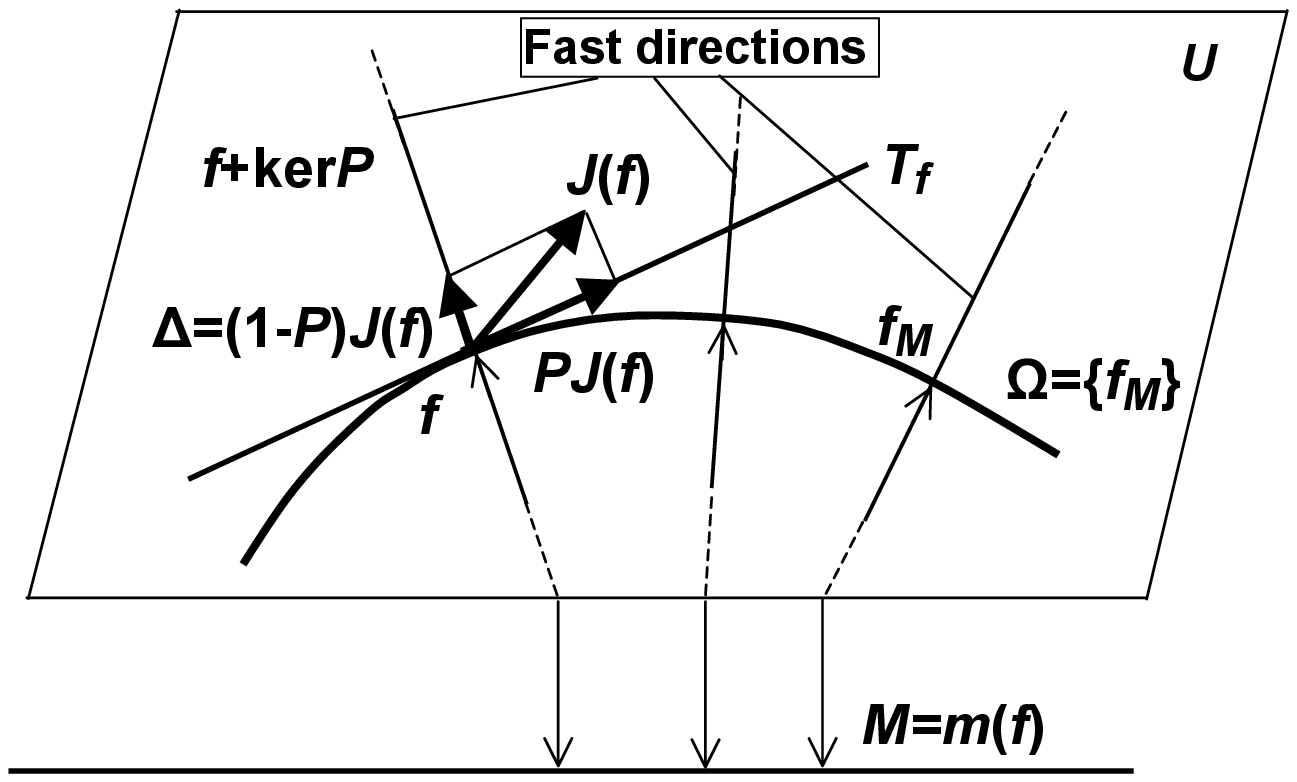}
\caption {\label{fig3QEHier}Parametrization by macroscopic
variables: linear (a), nonlinear (b) and layer-linear (c). Thin
arrows illustrate the bijection $M\leftrightarrow f_M$.  a) Moment
parametrization in fast-slow decomposition. Dashed lines -- the
plains of constant value of moments $M$. These plains coincide
with directions of fast motion in the moment approximation. b)
Non-linear macroscopic parametrization in fast-slow decomposition.
Dashed curves -- the surfaces of constant value of macroscopic variables $M$.
Plains of fast motion are tangent to these surfaces.  c)
Nonlinear, but layer--linear macroscopic parametrization in
fast-slow decomposition. The surfaces of constant value of macroscopic variables
$M$ (dashed lines) are plain, but the dependence $m(f)$ is
nonlinear. Plains of fast motion coincide with these plains.  }
\end{centering}
\end{figure}

For dissipative systems, we always keep in mind the following
picture (Fig.~\ref{fig1QEHier}b). The vector field $J(f)$
generates the motion on the phase space $U$: $\D f / \D t = J(f)$.
An ansatz manifold $\Omega$ is given, it is the current
approximation to the invariant manifold.

The projected vector field $PJ(f)$ belongs to the tangent space
$T_f$, and the equation $\D f / \D t = PJ(f) $ describes the motion
along the ansatz manifold $\Omega$ (if the initial state belongs to
$\Omega$).

The choice of the projector $P$ might be very important. There is
some ``duality" between accuracy of slow invariant manifold
approximation and restrictions on the projector choice. If
$\Omega$ is an exactly invariant manifold, then the vector field
$J(f)$ is tangent to $\Omega$, and all projectors give the same
result. If $\Omega$ gives a good smooth approximation for such an
invariant manifold, then the set of admissible projectors is
rather broad. On the other hand, there is the unique choice of the
projector applicable for every (arbitrary) ansatz $\Omega$
\cite{GK1,UNIMOLD}, any other choice leads to violation of the
Second Law in projected equations.

In the initial geometry  of the fast--slow decomposition
(Figs.~\ref{fig1QEHier}a, \ref{fig1QEHier}b) the ``slow variables"
(or ``macroscopic variables") are internal coordinates on the slow
manifold, or on its approximation $\Omega$. It is impossible, in
general, to define these macroscopic variables as functionals of
$f$ outside these manifolds. Moreover, this definition cannot be
unique.

The {\it moment parametrization} starts not from the manifold, but
from the macroscopic variables defined in the whole $U$
(Fig.~\ref{fig3QEHier}a), and for the given variables it is
necessary to find the corresponding slow manifold. Usually, these
slow variables are linear functions (functionals), for example,
hydrodynamic fields (density, momentum density, and pressure) are
moments of one-particle distribution functions $f(x,v)$. The
moment vector $M$ is the value of linear operator $m$: $M=m(f)$.
The moments value serve as internal coordinates on the
(hypothetic) approximate slow manifold $\Omega$. It means that
points of $\Omega$ are parameterized by $M$, $\Omega=\{f_M\}$, and
the consistency condition holds: $$m(f_M)=M.$$ In the example with
one-particle function $f$ and hydrodynamic fields $m(f)$ it means
that slow manifold consists of distribution $f(x,v)$ parameterized
by their hydrodynamic fields (as in the Chapman--Enskog method
\cite{Chapman}). For a given $\Omega=\{f_M\}$, the moment equation
has a very simple form:
\begin{equation} \label{Momsys}
{\D M \over \D t}= m(J(f_M)),
\end{equation}
and the correspondent equation for the projected motion on the
manifold $\Omega=\{f_M\}$ is
\begin{equation} \label{Momsyspro}
{\D f \over \D t}= (D_M f_M)m(J(f_M)),
\end{equation}
Where $D_M f_M$ is differential of the parametrization $M\mapsto
f_M$.

How to find a manifold $\Omega=\{f_M\}$ for a given moment
parametrization? A good initial approximation is the
quasiequilibrium (or MaxEnt) approximations. The basic idea is: in
the fast motion the entropy should increase, hence, the point of
entropy maximum on the plane of rapid motion is not far from the
slow manifold (Fig.~\ref{fig1QEHier}a). If our moments $M$ are
really slow variables, and don't change significantly in rapid
motion, then the manifold of conditional entropy maxima $f_M$:
\begin{equation}\label{emax}
S(f)\rightarrow\max,\; m(f)=M
\end{equation}
could serve as appropriate ansatz for slow manifold.

Most of the works on nonequilibrium thermodynamics deal with
quasiequilibrium approximations and corrections to them, or with
applications of these approximations (with or without
corrections). This viewpoint is not the only possible but it
proves very efficient for the construction of a variety of useful
models, approximations and equations, as well as methods to solve
them. From time to time it is discussed in the literature, who was
the first to introduce the quasiequilibrium approximations, and
how to interpret them. At least a part of the discussion is due to
a different role the quasiequilibrium plays in the
entropy-conserving and the dissipative dynamics. The very first
use of the entropy maximization dates back to the classical work
of G.\ W.\ Gibbs \cite{Gibb}, but it was first claimed for a
principle of informational statistical thermodynamics by E.\ T.\
Jaynes \cite{Janes1}. Probably the first explicit and systematic
use of quasiequilibria to derive dissipation from
entropy-conserving systems was undertaken by D.\ N.\ Zubarev.
Recent detailed exposition is given in \cite{Zubarev}. For
dissipative systems, the use of the quasiequilibrium to reduce
description can be traced to the works of H.\ Grad on the
Boltzmann equation \cite{Grad}. A review of the informational
statistical thermodynamics was presented in \cite{Garsia1}.  The
connection between entropy maximization and (nonlinear) Onsager
relations was also studied \cite{Nett,Orlov84}. The viewpoint  of
the present authors was influenced by the papers by L.\ I.\
Rozonoer and co-workers, in particular, \cite{KoRoz,Ko,Roz}. A
detailed exposition of the quasiequilibrium approximation for
Markov chains is given in the book \cite{G1} (Chap. 3, {\it
Quasiequilibrium and entropy maximum}, pp.\ 92-122), and for the
BBGKY hierarchy in the paper \cite{Kark}. The maximum entropy
principle was  applied to the description the universal dependence
the three-particle distribution function $F_3$ on the two-particle
distribution function $F_2$ in classical systems with binary
interactions \cite{BGKTMF}. For a discussion the quasiequilibrium
moment closure hierarchies for the Boltzmann equation \cite{Ko}
see the papers \cite{MBCh,MBChLANL,Lever}. A very general
discussion of the maximum entropy principle with applications to
dissipative kinetics is given in the review \cite{Bal}. Recently
the quasiequilibrium approximation with some further correction
was applied to description of rheology of polymer solutions
\cite{IKOePhA02,IKOePhA03} and of ferrofluids \cite{IlKr,IKar2}.
Quasiequilibrium approximations for quantum systems in the Wigner
representation \cite{WIG,CAL} was discussed very recently
\cite{Degon}.

Formally, for quasiequilibrium approximation the linearity of the
map $f \mapsto m(f)$ is not necessary, and the optimization
problem (\ref{emax}) could be studied for nonlinear conditions
$m(f)=M$ (Fig.~\ref{fig3QEHier}b). Nevertheless, the problem
(\ref{emax}) with nonlinear conditions loose many important
properties caused by concavity of $S$. The technical compromise is
the problem with nonlinear map $m$, but linear restrictions
$m(f)=M$. It is possible when preimages of points for the map $m$
are plain (Fig.~\ref{fig3QEHier}c). Such a ``layer--linear"
approximation for a generic smooth map $m_0: f \mapsto M$ could be
created as follows. Let $\Omega_0$ be a smooth submanifold in $U$.
In some vicinity of $\Omega_0$ we define a map $m$
\begin{equation}\label{layerlinearization}
m(f)=m_0(f_0) \; \; \mbox{if} \; \; (Dm_0)_{f_0} (f-f_0)=0,
\end{equation}
where $f_0$ are points from $\Omega_0$ and $(Dm_0)_{f_0}$ is the
differential of $m_0$ at the point $f_0$. This definition means
that $m(f)=m_0(f_0)$ if  $m_0(f)$ coincides with $m_0(f_0)$ in
linear approximation.

The layer--linear parametrization was introduced in Ref.
\cite{GKMod} for construction of generalized model equations for
the Boltzmann kinetics.

Eq.~(\ref{layerlinearization}) defines a smooth layer--linear map
$m$ in a vicinity of $\Omega_0$ under some general transversality
condition.

Let an initial approximation $\Omega_0$ for the slow manifold be
unsatisfactory. Two basic ways for improvement are: (i) manifold
correction and (ii) manifold extension. On the first way we should
find a shifted manifold that is better approximate slow invariant
manifold. The list of macroscopic variables remains the same. On
the second way we extend the list of macroscopic variables, and,
hence, extend the manifold $\Omega$. The Chapman--Enskog method
\cite{Chapman} gives the example of manifold correction in the
form of Taylor series expansion, the direct Newton method could
give better results
\cite{GK1,GKTTSP94,CMIM,GKSpri,GKIOeNONNEWT2001}. The second way
is the essence of EIT -- extended irreversible thermodynamics
\cite{EIT}. This paper is focused on the manifold extensions also.

Usually moments are graduated in a natural order, by degree of
polynomials: concentration (zero order of velocity), average
momentum density (first order), kinetic energy (second order),
stress tensor (second order), heat flux (third order), etc. Normal
logic of EIT is extension of the list of variables by addition of
the next irreducible moment tensor. But there is another logic. In
general, for the set of moments $M$ that parametrizes $\Omega_0$ a
time derivative is a known function of $f$: $\D M / \D t =F_M
(f)$. We propose to construct new moments from $F_M (f)$. It
allows to achieve a best possible approximation for $\D M / \D t$
through extended variables. For this nonlinear variables we use
the layer--linear approximation (\ref{layerlinearization}), ass
well, as a layer--quadratic approximation for the entropy. This
(layer) linearization of the problem near current approximation
follows lessons of the Newton method.

It should be stressed that ``layer--linear" does not mean
``linear", and the modified choice of new variables implies no
additional restrictions, but it pretends to be just a more direct
way to dynamic invariance. Below this approach is demonstrated on
the Boltzmann equation.

\section{Difficulties of classical methods of the Boltzmann
equation theory}

The Boltzmann equation remains the most inspiring source for the
model reduction problems. The first systematic and (at least
partially) successful method of constructing invariant manifolds
for dissipative systems was the celebrated {\it Chapman--Enskog
method} \cite{Chapman} for the Boltzmann kinetic equation. The
main difficulty of the Chapman--Enskog method \cite{Chapman} are
``nonphysical" properties of high-order approximations. This was
stated by a number of authors and was discussed in detail in
\cite{Cercignani}. In particular, as it was noted in \cite{Bob},
the Burnett approximation results in a short-wave instability of
the acoustic spectra. This fact contradicts the $H$-theorem (cf.
in \cite{Bob}). The Hilbert expansion contains secular terms
\cite{Cercignani}. The latter contradicts the $H$-theorem.
\par
The other difficulties of both  of  these  methods  are:  the
restriction upon the choice of the initial approximation  (the local
equilibrium approximation), the requirement for a small parameter,
and the usage  of  slowly   converging   Taylor expansion.   These
difficulties never allow a direct transfer  of these  methods  on
essentially nonequilibrium situations.
\par
The  main  difficulty  of  the  Grad  method   \cite{Grad}   is
the uncontrollability of the chosen approximation. An extension of
the list of moments can result in a certain success, but it  can
also give nothing. Difficulties of moment expansion in the problems
of shock waves and sound propagation are discussed in
\cite{Cercignani}.
\par
Many attempts were made to refine these methods. For the
Chapman--Enskog and  Hilbert methods  these attempts  are based in
general on some better rearrangement of expansions  (e.g.
neglecting high-order  derivatives  \cite{Cercignani}, reexpanding
\cite{Cercignani},  Pade approximations and partial summing
\cite{MBCh,GKJETP91,Slem2,KGAnPh2002}, etc.). This type of work
with formal series is wide spread  in  physics. Sometimes  the
results are surprisingly good -- from the renormalization theory
in quantum fields to the Percus-Yevick equation and the
ring-operator in statistical mechanics. However,  one  should
realize  that success cannot be guaranteed. Moreover,
rearrangements never remove the restriction  upon  the  choice  of
the  initial  local equilibrium approximation.
\par
Attempts  to  improve  the   Grad   method   are   based   on
quasiequilibrium approximations \cite{KoRoz,Ko}. It was found in
\cite{Ko}  that the Grad  distributions  are  linearized versions
of appropriate quasiequilibrium approximations (see also
\cite{MBCh,MBChLANL,Lever}). A method which treats fluxes (e.g.
moments with respect to collision integrals) as  independent
variables in a quasiequilibrium description was introduced in
\cite{MBCh,MBChLANL,Karlin1,GKPRE96}, and will be discussed later.

The important feature of quasiequilibrium approximations is that
they are always thermodynamic, i.e. they are consistent  with the
$H$-theorem  by  construction.

\section{Triangle Entropy Method}

In the present subsection, which is of introductory character, we
shall refer, to be specific, to the Boltzmann kinetic equation for
a one-component gas  whose state (in the microscopic sense) is
described by the one-particle distribution function
$f({\vv},{\xx},t)$ depending on the velocity vector ${\vv}=\left\{
v_{k}\right\} ^{3}_{k=1}$, the spatial position $\xx=\left\{
x_{k}\right\} ^{3}_{k=1}$ and time $t$. The Boltzmann equation
describes the evolution of $f$ and in the absence of external
forces is
\begin{equation}
\partial_{t} f+v_{k} \partial_{k}f=Q(f,f),
\label{Tri1.1}\end{equation} \noindent where $\partial_{t} \equiv
\partial  /\partial t$ is the time partial  derivative, $
\partial_{k}\equiv \partial / \partial x_{k}$  is  partial
derivative with respect to $k$-th component of ${\xx}$, summation in
two  repeating  indices  is  assumed, and $Q(f,f)$  is  the
collision integral (its concrete form is of no importance right now,
just note that it is functional-integral operator  quadratic with
respect to $f$).
\par
The Boltzmann equation possesses two properties principal for the
subsequent reasoning.
\par
1. There exist five functions $\psi _{\alpha }({\vv})$ (additive
collision invariants),
$$1, {\vv}, {v}^{2}$$ such  that  for  any  their  linear combination with coefficients depending
on ${\xx}, t$ and for  arbitrary $f$ the following equality  is
true:
\begin{equation}
\int \sum_ {\alpha =1}^5 a_{\alpha }({\xx},t)\psi _{\alpha
}({\vv})Q(f,f) \,\D {\vv}=0, \label{Tri1.2}
\end{equation}
provided the integrals exist.

2. The equation (\ref{Tri1.1})  possesses  global  Lyapunov
functional: the $H$-function,
\begin{equation}
H(t)\equiv H[f]=\int f({\vv},{\xx},t)\ln f({\vv},{\xx},t) \,\D {\vv}
\,\D {\xx}, \label{Tri1.3}\end{equation}

\noindent the  derivative  of  which  by  virtue  of  the  equation
(\ref{Tri1.1})  is non-positive under appropriate boundary
conditions:
\begin{equation}
\D H(t)/ \D t\leq 0. \label{Tri1.4}\end{equation}

Grad's method \cite{Grad} and its variants construct closed systems
of equations for macroscopic variables when the latter are
represented by moments (or, more general, linear functionals) of the
distribution function $f$ (hence their  alternative  name is the
``moment methods"). The entropy maximum method for the Boltzmann
equation is of particular importance for the subsequent reasoning.
It consists in the following. A finite set of moments describing the
macroscopic state is chosen. The distribution function of the
quasiequilibrium state  under given values  of the chosen moments is
determined, i.e. the problem is solved
\begin{equation}
H[f] \rightarrow  \min,\hbox{  for }  \hat{M}_i [f]=M_i,  \ \
i=1,\ldots,k, \label{Tri1.5}\end{equation} \noindent where
$\hat{M}_{i}[f]$ are linear functionals with respect to $f$; $M_{i}$
are  the corresponding values of chosen set of $k$ macroscopic
variables. The quasiequilibrium distribution function
$f^{*}({\vv},M({\xx},t))$, $M=\left\{ M_{1},\ldots ,M_{k}\right\}$,
parametrically depends on $M_{i}$, its dependence on space ${\xx}$
and on time $t$ being represented only by $M({\xx},t)$. Then the
obtained $f^{*}$ is substituted into  the Boltzmann equation
(\ref{Tri1.1}), and operators $\hat{M}_i$ are applied on the latter
formal expression.
\par
In the result we have closed systems of equations with  respect to
$M_{i}({\xx},t)$, $i=1,\ldots,k$:
\begin{equation}
\partial_{t} M_{i}+\hat{M}_{i}[{\vv}_{k} \partial_{k}
f^{*}({\vv},M)]=\hat{M}_{i}[Q(f^{*}({\vv},M),f^{*}({\vv},M))] .
\label{Tri1.6}\end{equation}

The following heuristic explanation can be given  to the entropy
method. A state of the gas can be described  by  a  finite set of
moments on some time scale $\theta $ only if all the other moments
(``fast") relax  on a shorter time scale time $\tau, \tau<<\theta $,
to their  values  determined  by the chosen set of ``slow" moments,
while the slow ones almost do not change appreciably on the time
scale $\tau$. In the process of the fast relaxation the $H$-function
decreases,  and  in the end  of  this fast relaxation process a
quasiequilibrium state  sets in with  the distribution function
being the solution  of the problem (\ref{Tri1.5}).  Then ``slow"
moments relax to the equilibrium state by virtue of (\ref{Tri1.6}).

The entropy  method  has  a  number  of  advantages  in comparison
with the classical Grad's method. First, being  not necessarily
restricted to any specific  system   of orthogonal polynomials, and
leading to solving an optimization problem, it is more convenient
from the technical point of view. Second, and  ever more important,
the   resulting quasiequilibrium $H$-function,
$H^{*}(M)=H[f^{*}({\vv},M)]$, decreases due of the moment equations
(\ref{Tri1.6}).

Let us  note  one  common  disadvantage  of  all  the  moment
methods,   and,   in   particular,   of the   entropy    method.
Macroscopic parameters, for which  these  methods  enable  to obtain
closed systems, must be moments of the distribution function. On
the  other  hand,  it  is   easy   to find examples when the
interesting macroscopic parameters  are  nonlinear functionals   of
the distribution function. In the case of the one-component gas
these are  the  integrals of velocity polynomials with respect to
the collision integral $Q(f,f)$ of (\ref{Tri1.1}) (scattering rates
of moments). For chemically reacting mixtures these are  the
reaction rates, and so on. If the characteristic relaxation  time of
such nonlinear macroscopic parameters  is comparable  with that of
the  ``slow" moments, then they should be also included into the
list of  ``slow" variables on the same footing.

In this paper for  constructing  closed  systems  of equations for
non-linear  (in a general case) macroscopic variables the {\it
triangle entropy method} is used. Let us outline the scheme of this
method.

Let a set of macroscopic variables be chosen: linear functionals
$\hat{M}[f]$ and nonlinear functionals  (in a general case)
$\hat{N}[f]$: $$ \hat{M}[f]=\left\{ \hat{M}_{1}[f],\ldots
,\hat{M}_{k}[f]\right\}, \: \:  \hat{N}[f]=\left\{ \hat{N}_{1}[f],
\ldots ,\hat{N}_{l}[f]\right\}. $$ Then, just as for the problem
(\ref{Tri1.5}), the first quasiequilibrium approximation is
constructed under fixed values of the linear macroscopic parameters
$M$:
\begin{equation}
H[f] \rightarrow  \min \hbox{  for } \hat{M}_{i}[f]=M_i,\ i=1,\ldots
,k, \label{Tri1.7}\end{equation}

\noindent and the resulting distribution function is
$f^{*}({\vv},M)$.  After  that, we seek the true quasiequilibrium
distribution function in the form,
\begin{equation}
f=f^{*}(1+\varphi ),\label{Tri1.8}
\end{equation}

\noindent where $\varphi $  is  a  deviation  from   the   first
quasiequilibrium approximation. In order to determine $\varphi$,
the  second  quasiequilibrium approximation is constructed. Let us
denote $\Delta H[f^{*},\varphi ]$ as  the  quadratic term in the
expansion of the $H$-function into powers of $\varphi $  in  the
neighbourhood   of   the   first   quasiequilibrium   state
$f^{*}$. The distribution function   of    the    second
quasiequilibrium approximation is the solution to the problem,
\begin{eqnarray}
&&\Delta H [f^*,\varphi] \rightarrow \hbox{min   for} \nonumber \\
&& \hat{M}_i[f^*\varphi] = 0,\ \ i=1,\ldots, k, \nonumber \\
&&\Delta \hat{N}_{j} [f^*,\varphi] = \Delta N_j,\ \ j=1,\ldots ,l,
\label{Tri1.9}
\end{eqnarray}

\noindent where $\Delta \hat{N}_{j}$ are linear  operators
characterizing the linear  with respect to $\varphi $ deviation  of
(nonlinear)  macroscopic parameters $N_{j}$  from their values,
$N^{*}_{j}=\hat{N}_{j}[f^{*}]$, in the first quasiequilibrium
state. Note the importance of the homogeneous constraints
$\hat{M}_i[f^*\varphi] = 0$ in the problem (\ref{Tri1.9}).
Physically, it means that the variables $\Delta N_j$ are ``slow" in
the same sense, as the variables $M_i$, at least in the small
neighborhood of the first quasiequilibrium $f^*$.  The obtained
distribution function,
\begin{equation}
f=f^{*}({\vv},M)(1+\varphi ^{**}({\vv},M,\Delta N))
\label{Tri1.10}\end{equation}

\noindent is used  to  construct  the  closed  system  of  equations
for  the macroparameters $M$, and $\Delta  N$. Because the
functional in the problem (\ref{Tri1.9}) is quadratic, and all
constraints in this problem are linear, it is always explicitly
solvable.

Further in this section some examples of using  the  triangle
entropy method for the one-component gas are considered.
Applications to chemically reacting mixtures were discussed in
\cite{Karlin1}.

\section{Linear macroscopic variables}

Let us consider the simplest example of using the  triangle entropy
method, when all the macroscopic variables of the first and of  the
second quasiequilibrium states are the  moments of the distribution
function.

\subsection{Quasiequilibrium projector}

Let $\mu _{1}({\vv}),\ldots ,\mu _{k}({\vv})$ be the microscopic
densities of  the  moments
$$M_{1}({\xx},t),..., M_{k}({\xx},t)$$  which determine  the  first quasiequilibrium state,
\begin{equation}
M_{i}({\xx},t)=\int \mu _{i}({\vv})f({\vv},{\xx},t) \,\D {\vv},
\label{Tri2.1}\end{equation} \noindent and let $\nu
_{1}({\vv}),\ldots ,\nu _{l}({\vv})$ be the microscopic densities of
the moments $$N_{1}({\xx},t),..., N_{l}({\xx},t)$$ determining
together  with  (\ref{Tri1.1}) the second quasiequilibrium state,
\begin{equation}
N_{i}({\xx},t)=\int \nu _{i}({\vv})f({\vv},{\xx},t) \,\D {\vv} .
\label{Tri2.2}\end{equation} \noindent The choice of the set of the
moments of the first and  second quasiequilibrium approximations
depends on a specific problem.  Further on  we assume that the
microscopic density $\mu \equiv 1$ corresponding to the
normalization condition is always included in the list of
microscopic densities of  the moments  of  the  first
quasiequilibrium state. The distribution function of the first
quasiequilibrium state results from solving the optimization
problem,
\begin{equation}
H[f]=\int f({\vv}) \ln f({\vv})  \,\D {\vv}\rightarrow  \min
\label{Tri2.3}\end{equation} for $$ \int \mu _{i}({\vv})f({\vv})
\,\D {\vv}=M_{i}, i=1,\ldots, k. $$

\noindent Let us denote by $M=\left\{ M_{1},\ldots ,M_{k}\right\} $
the   moments  of the first quasiequilibrium state, and by
$f^{*}({\vv},M)$ let us denote the solution of the problem
(\ref{Tri2.3}).
\par
The distribution function of the second  quasiequilibrium  state is
sought in the form,
\begin{equation}
f=f^{*}({\vv},M)(1+\varphi ). \label{Tri2.4}\end{equation} \noindent
Expanding the $H$-function (\ref{Tri1.3}) in the neighbourhood  of
$f^{*}({\vv},M)$  into powers of $\varphi $ to second order we
obtain,
\begin{eqnarray}
\Delta H({\xx},t)\equiv \Delta H[f^{*},\varphi ]&=&H^{*}(M)+ \int
f^{*}({\vv},M)\ln f^{*}({\vv},M)\varphi ({\vv}) \,\D {\vv} \nonumber
\\ &&+{1\over 2} \int f^{*}({\vv},M)\varphi ^{2}({\vv}) \,\D {\vv},
\label{Tri2.5} \end{eqnarray} \noindent where
$H^{*}(M)=H[f^{*}({\vv},M)]$ is the value of the $H$-function  in
the  first quasiequilibrium state.

When searching for the second quasiequilibrium state, it  is
necessary that the true values of the moments $M$ coincide with
their values in the first quasiequilibrium state, i.e.,
\begin{eqnarray}
M_{i}&=&\int \mu _{i}({\vv})f^{*}({\vv},M)(1+\varphi ({\vv})) \,\D
{\vv} \nonumber \\ &=& \int \mu _{i}({\vv})f^{*}({\vv},M) \,\D
{\vv}=M^{*}_{i}, \; i=1,\ldots ,k. \label{Tri2.6}
\end{eqnarray}
In other words, the set of the homogeneous conditions on $\varphi $
in  the problem (\ref{Tri1.9}),
\begin{equation}
\int \mu _{i}({\vv})f^{*}({\vv},M)\varphi ({\vv}) \,\D {\vv}=0,
i=1,\ldots ,k, \label{Tri2.7}\end{equation} \noindent ensures a
shift (change) of the first quasiequilibrium state only due to the
new moments $N_{1},\ldots ,N_{l}$.  In order to take this condition
into account automatically, let us introduce the following structure
of a Hilbert space:
\begin{enumerate}
\item{Define the scalar product
\begin{equation}
(\psi _{1},\psi _{2})=\int f^{*}({\vv},M)\psi _{1}({\vv})\psi
_{2}({\vv})  \,\D {\vv}. \label{Tri2.8}\end{equation} } \item{Let
$E_{\mu}$ be the linear hull of the set  of  moment densities
 $$\{ \mu _{1}({\vv}),\ldots,\mu _{k}({\vv}) \}.$$ Let us construct a basis of
 $E_{\mu}$ $\{ e_{1}({\vv}),\ldots,e_{r}({\vv} )\}$ that is orthonormal in the sense of the scalar product (\ref{Tri2.8}):
\begin{equation}
(e_{i},e_{j})=\delta _{ij}, \label{Tri2.9}\end{equation} \noindent
$i,j=1,\ldots ,r; \delta _{ij}$ is the Kronecker delta.}
\item{Define a projector $\hat{P}^{*}$ on the first
quasiequilibrium state,
\begin{equation}
\hat{P}^{*}\psi  =\sum_{i=1}^r e_{i}(e_{i},\psi ).
\label{Tri2.10}\end{equation} The projector $\hat{P}^{*}$ is
orthogonal: for any pair of functions $\psi _{1}, \psi _{2}$,
\begin{equation}
(\hat{P}^{*}\psi _{1},(\hat{1}-\hat{P}^{*})\psi _{2})=0,
\label{Tri2.11}\end{equation} \noindent where \^{1} is the unit
operator. Then the condition (\ref{Tri2.7}) amounts to
\begin{equation}
\hat{P}^{*}\varphi =0, \label{Tri2.12}\end{equation} \noindent and
the expression for the quadratic part of the $H$-function
(\ref{Tri2.5}) takes  the form,
\begin{equation}
\Delta H[f^{*},\varphi ]=H^{*}(M)+(\ln f^{*},\varphi )+(1/2)(\varphi
,\varphi ). \label{Tri2.13}
\end{equation}
}
\end{enumerate}

Now, let us note that the function $\ln  f^{*}$  is  invariant with
respect to the action of the projector $\hat{P}^{*}$:
\begin{equation}
\hat{P}^{*} \ln f^{*} =\ln f^{*}. \label{Tri2.14}
\end{equation}
\par
\noindent This follows directly  from the solution  of  the  problem
(\ref{Tri2.3}) using of the method of Lagrange multipliers:

$$ f^{*} =\exp \sum_{i=1}^k \lambda_{i} (M) \mu_{i} ({\vv}), $$

\noindent where $\lambda_{i}(M)$ are Lagrange multipliers. Thus, if
the condition (\ref{Tri2.12}) is satisfied, then from
(\ref{Tri2.11}) and (\ref{Tri2.14}) it follows that

$$ (\ln  f^{*},\varphi )=(\hat{P}^{*} \ln f^{*},(\hat{1}-\hat{P}^{*})\varphi )=0. $$

Condition (\ref{Tri2.12}) is satisfied automatically, if   $\Delta
N_{i}$ are taken as follows:
\begin{equation}
\Delta N_{i}=((\hat{1}-\hat{P}^{*})\nu _{i},\varphi ), i=1,\ldots ,
l.  \label{Tri2.15}
\end{equation}

Thus,   the   problem   (\ref{Tri1.9})   of   finding    the
second quasiequilibrium state reduces to
\begin{eqnarray}
\Delta H[f^{*},\varphi]-H^{*}(M)=(1/2)(\varphi ,\varphi )\rightarrow  \min \hbox{ for } \nonumber \\
((\hat{1}-\hat{P}^{*})\nu _{i},\varphi )=\Delta N_{i},\ \
i=1,\ldots,l. \label{Tri2.16}
\end{eqnarray}

Note that it is  not  ultimatively  necessary  to  introduce  the
structure of the Hilbert  space. Moreover that may  be impossible,
since the ``distribution function" and the ``microscopic  moment
densities" are, strictly  speaking, elements  of  different
(conjugate one to another) spaces, which may be not reflexive.
However, in the examples considered below the mentioned difference
is  not manifested.

In the remainder of this section we demonstrate how the triangle
entropy method is related to Grad's moment method.

\subsection{Ten-moment Grad approximation.}

Let  us  take  the  five  additive collision invariants as moment
densities of  the first quasiequilibrium  state:
\begin{equation}
\mu _{0}=1; \: \mu _{k}=v_{k} \, (k=1,2,3); \: \mu _{4}={mv^{2}
\over 2}, \label{Tri2.17}\end{equation} \noindent where $v_{k}$ are
Cartesian components of the velocity, and $m$ is particle's mass.
Then the solution to the  problem  (\ref{Tri2.3})  is  the  local
Maxwell distribution function $f^{(0)}({\vv},{\xx},t)$:
\begin{equation}
f^{(0)}=n({\xx},t) \left({2\pi k_{\rm B}T({\xx},t) \over m}
\right)^{-3/2} \exp \left\{ - {m({\vv}-{\uu}({\xx},t))^{2}\over
2k_{\rm B}T({\xx},t)} \right\}, \label{Tri2.18}\end{equation}
\noindent where

\noindent$n({\xx},t)=\int f({\vv}) \,\D {\vv}$ is   local  number
density,

\noindent${\uu}({\xx},t)=n^{-1}({\xx},t) \int f({\vv}){\vv}  \,\D
{\vv}$    is  the  local    flow density,

\noindent$T({\xx},t)=\frac{m}{3k_{\rm B}}n^{-1}({\xx},t)\int
f({\vv})({\vv}-{\uu}({\xx},t))^{2}
 \,\D {\vv}$
 is the local temperature,

\noindent$k_{\rm B}$ is  the  Boltzmann constant.

Orthonormalization of the set of moment densities  (\ref{Tri2.17})
with the weight (\ref{Tri2.18}) gives one of the possible
orthonormal basis
\begin{eqnarray}
e_{0}&=&{5k_{\rm B}T-m({\vv}-{\uu})^{2} \over (10n)^{1/2}k_{\rm B}T}, \nonumber \\
 e_{k}&=&{m^{1/2}(v_{k}-u_{k}) \over (nk_{\rm B}T)^{1/2}}, k=1,2,3, \label{Tri2.19} \\
e_{4}&=&{m({\vv}-{\uu})^{2} \over (15n)^{1/2}k_{\rm B}T}. \nonumber
\end{eqnarray}

For the moment densities of the second quasiequilibrium state let us
take,
\begin{equation}
\nu _{ik}=mv_{i}v_{k}, \: i,k=1,2,3. \label{Tri2.20} \end{equation}
\noindent Then
\begin{equation}
(\hat{1}-\hat{P}^{(0)})\nu _{ik}=m(v_{i}-u_{i})(v_{k}-u_{k})-{1\over
3}\delta _{ik}m({\vv}-{\uu})^{2}, \label{Tri2.21}\end{equation}
\noindent and, since $((\hat{1}-\hat{P}^{(0)})\nu
_{ik},(\hat{1}-\hat{P}^{(0)})\nu _{ls})= (\delta _{il}\delta
_{ks}+\delta _{kl}\delta _{is})Pk_{\rm B}T/m$,  where $P=nk_{\rm
B}T$ is  the  pressure,  and $\sigma
_{ik}=(f,(\hat{1}-\hat{P}^{(0)})\nu _{ik})$   is the traceless part
of the stress tensor, then from (\ref{Tri2.4}), (\ref{Tri2.17}),
(\ref{Tri2.18}),  (\ref{Tri2.21}) we obtain the distribution
function of the second quasiequilibrium state in the form
\begin{equation}
f=f^{(0)} \left( 1+{\sigma _{ik}m\over 2Pk_{\rm B}T} \left[
(v_{i}-u_{i})(v_{k}-u_{k})-{1\over 3} \delta _{ik}({\vv}-{\uu})^{2}
\right] \right) \label{Tri2.22}
\end{equation}
\noindent This is precisely  the  distribution  function  of  the
ten-moment Grad approximation (let us recall  that  here summation
in two repeated indices is assumed).

\subsection{Thirteen-moment Grad approximation}

In addition  to  (\ref{Tri2.17}), (\ref{Tri2.20}), let us extend the
list of  moment  densities  of the second quasiequilibrium state
with the functions
\begin{equation}
\xi _{i}={mv_{i}v^{2}\over 2}, \: i=1,2,3.
\label{Tri2.23}\end{equation}

\noindent The corresponding orthogonal complements to the projection
on the first quasiequilibrium state are

\begin{equation}
(\hat{1}-\hat{P}^{(0)})\xi _{i}={m\over 2}(v_{i}-u_{i}) \left(
({\vv}-{\uu})^{2}-{5k_{\rm B}T\over m} \right).
\label{Tri2.24}\end{equation}

\noindent The moments corresponding to  the  densities
$(\hat{1}-\hat{P}^{(0)})\xi _{i}$  are  the components of the heat
flux vector $q_{i}$:

\begin{equation}
q_{i}=(\varphi ,(\hat{1}-\hat{P}^{(0)})\xi _{i}).
\label{Tri2.25}\end{equation}

\noindent

Since $$ ((\hat{1}-\hat{P}^{(0)})\xi _{i},(\hat{1}-\hat{P}^{(0)})\nu
_{lk})=0 ,$$

\noindent for any $i, k, l$, then the constraints

$$ ((\hat{1}-\hat{P}^{(0)})\nu _{lk},\varphi )=\sigma _{lk}, ((\hat{1}-\hat{P}^{(0)})\xi _{i},\varphi
)=q_{i} $$

\noindent in the problem (\ref{Tri2.16}) are independent,  and
Lagrange  multipliers corresponding to $\xi _{i}$ are
\begin{equation}
{1\over 5n} \left({k_{\rm B}T\over m} \right)^{2} q_{i}.
\label{Tri2.26}\end{equation}

Finally, taking into account (\ref{Tri2.17}), (\ref{Tri2.22}),
(\ref{Tri2.24}), (\ref{Tri2.26}), we find the distribution function
of the second quasiequilibrium state in the form
\begin{eqnarray}
f=f^{(0)} \left(1+{\sigma _{ik}m \over 2Pk_{\rm B}T}
\left((v_{i}-u_{i})(v_{k}-u_{k})-{1\over 3}\delta
_{ik}({\vv}-{\uu})^{2} \right) \right. \nonumber \\ \left.+
{q_{i}m\over Pk_{\rm B}T}(v_{i}-u_{i}) \left(
{m({\vv}-{\uu})^{2}\over 5k_{\rm B}T} - 1 \right) \right) ,
\label{Tri2.27} \end{eqnarray}

\noindent which  coincides  with  the  thirteen-moment   Grad
distribution function \cite{Grad}.

Let us remark on the thirteen-moment approximation. From
(\ref{Tri2.27}) it follows that for large enough negative  values
of $(v_{i}-u_{i})$  the thirteen-moment distribution function
becomes negative. This peculiarity of the thirteen-moment
approximation is due to the fact that the moment density $\xi _{i}$
is odd-order polynomial of $v_{i}$. In order to eliminate this
difficulty, one may consider from the very beginning that in a
finite volume the  square  of  velocity  of  a particle does not
exceed a certain value $v^{2}_{\max}$, which is finite owing to the
finiteness of the total energy, and $q_{i}$ is such that when
changing to infinite volume $q_{i}\rightarrow 0,
v^{2}_{\max}\rightarrow \infty $ and
$q_{i}(v_{i}-u_{i})({\vv}-{\uu})^{2}$  remains finite.

On the other hand, the solution to the optimization problem
(\ref{Tri1.5}) does not exist (is not normalizable), if the
highest-order velocity polynomial is odd, as it is for the full
13-moment quasiequilibrium.

Approximation (\ref{Tri2.22}) yields $\Delta H$ (\ref{Tri2.13}) as
follows:
\begin{equation}
\Delta H=H^{(0)}+n{\sigma _{ik}\sigma _{ik}\over 4P^{2}},
\label{Tri2.28}\end{equation}

\noindent while $\Delta H$  corresponding to (\ref{Tri2.27}) is,
\begin{equation}
\Delta H=H^{(0)}+ n{\sigma _{ik}\sigma _{ik}\over 4P^{2}} +
n{q_{k}q_{k}\rho \over 5P^{3}}, \label{Tri2.29}\end{equation}

\noindent where $\rho =mn$, and $H^{(0)}$ is the local equilibrium
value of the $H$-function
\begin{equation}
H^{(0)}= {5\over 2} n \ln n - {3\over 2} n \ln P - {3\over 2}n
\left( 1+\ln {2\pi \over m} \right). \label{Tri2.30}\end{equation}

These expressions coincide with the corresponding  expansions of the
quasiequilibrium $H$-functions obtained by the  entropy  method, if
microscopic moment densities of the first  quasiequilibrium
approximation are chosen as  $1, v_{i}$, and $v_{i}v_{j}$, or as $1,
v_{i}$, $v_{i}v_{j}$, and $v_{i}v^{2}$. As it was noted in
\cite{Ko}, they differs from  the $H$-functions obtained  by  the
Grad  method (without  the   maximum   entropy hypothesis), and in
contrast to the latter they give proper entropy balance equations.

The transition to the closed system of equations for the  moments of
the  first  and of the second quasiequilibrium  approximations  is
accomplished by proceeding from  the  chain  of  the  Maxwell moment
equations,  which  is  equivalent  to  the   Boltzmann   equation.
Substituting $f$ in the form of $f^{(0)}(1+\varphi )$ into  equation
(\ref{Tri1.1}), and multiplying  by $\mu _{i}({\vv})$, and
integrating over ${\vv}$, we obtain
\begin{eqnarray}
&&\partial_{t}(1,\hat{P}^{(0)} \mu_{i}({\vv}))+\partial _{t}(\varphi
({\vv}),\mu _{i}({\vv}))+\partial _{k}(v_{k}\varphi ({\vv}),\mu
_{i}({\vv})) \nonumber \\ && + \partial _{k}(v_{k},\mu _{i}({\vv}))
=M_{Q} [\mu _{i},\varphi ]. \label{Tri2.31} \end{eqnarray} \noindent

\noindent Here

$$ M_{Q}[\mu _{i},\varphi ]= \int Q(f^{(0)}(1+\varphi ),f^{(0)}(1+\varphi ))\mu _{i}({\vv}) \,\D
{\vv} $$

\noindent is a ``moment" (corresponding to the microscopic  density)
$\mu _{i}({\vv})$ with respect to the  collision  integral  (further
we term  $M_{Q}$ the collision moment or the scattering rate). Now,
if one uses $f$ given by equations (\ref{Tri2.22}), and
(\ref{Tri2.27}) as a closure assumption, then the system
(\ref{Tri2.31}) gives the ten- and thirteen-moment  Grad equations,
respectively, whereas only linear terms in $\varphi$ should be kept
when calculating $M_{Q}$.

Let us note some limitations of truncating the moment hierarchy
(\ref{Tri2.31}) by  means  of  the quasiequilibrium distribution
functions (\ref{Tri2.22}) and (\ref{Tri2.27}) (or for any other
closure which depends on the moments of the distribution functions
only).  When  such  closure is used, it is assumed implicitly that
the  scattering rates in the right hand side of (\ref{Tri2.31})
``rapidly" relax to their values determined by ``slow"
(quasiequilibrium) moments. Scattering rates are, generally
speaking, independent  variables. This peculiarity of the chain
(\ref{Tri2.31}), resulting from the nonlinear character of the
Boltzmann equation, distinct it essentially from the other
hierarchy equations of  statistical mechanics  (for example, from
the BBGKY chain which follows from the linear Liouville equation).
Thus, equations (\ref{Tri2.31}) are not closed twice: into the
left hand side of the equation for the $i$-th moment enters the
($i+1$)-th moment, and the right hand side contains additional
variables -- scattering rates.

A consequent way of closure of (\ref{Tri2.31}) should address both
sets of  variables  (moments  and scattering rates)  as  independent
variables. The triangle entropy method enables to do this.

\section{Transport equations for scattering rates in the
neighbourhood of local equilibrium.  Second and mixed hydrodynamic
chains}

In this section we derive equations of motion for the scattering
rates. It  proves  convenient  to use the following form of the
collision integral $Q(f,f)$:
\begin{equation}
Q(f,f)({\vv})=\int w({\vv'_1, v'} | {\vv,v_{1}}) \left(
f({\vv'})f({\vv'_1})-f({\vv})f({\vv}_{1}) \right) \,\D {\vv'} \,\D
{\vv'_1} \,\D {\vv_{1}}, \label{Tri3.1}\end{equation} where ${\vv}$
and ${\vv}_{1}$ are  velocities  of the two colliding particles
before the collision, ${\vv'}$ and ${\vv'_1}$ are  their velocities
after  the collision, $w$ is a kernel responsible for the
post-collision relations ${\vv'}({\vv},{\vv}_{1})$ and ${\vv'_1}
({\vv},{\vv}_{1})$, momentum and energy conservation laws  are
taken into account in $w$ by means of  corresponding $\delta
$-functions.  The kernel $w$  has the following symmetry property
with respect to   its arguments:
\begin{equation}
w({\vv'_1},{\vv'} |{\vv},{\vv_1})=w({\vv'_1},{\vv'}|
{\vv_1},{\vv})=w({\vv'},{\vv'_1} \mid
{\vv_1},{\vv})=w({\vv},{\vv_1}\mid {\vv'},{\vv'_1}).
\label{Tri3.2}\end{equation}

Let $\mu ({\vv})$  be  the microscopic  density  of  a  moment  $M$.
The corresponding scattering rate  $M_{Q}[f,\mu ]$ is defined as
follows:
\begin{equation}
M_{Q}[f,\mu ]= \int Q(f,f)({\vv})\mu ({\vv}) \,\D {\vv}.
\label{Tri3.3}\end{equation}

First, we should  obtain  transport  equations  for scattering rates
(\ref{Tri3.3}), analogous  to the moment's   transport equations.
Let us  restrict ourselves to the case when $f$ is represented in
the form,
\begin{equation}
f=f^{(0)}(1+\varphi ), \label{Tri3.4}\end{equation}

\noindent where $f^{(0)}$ is local Maxwell distribution function
(\ref{Tri2.18}),  and all the quadratic with respect to $\varphi $
terms will be neglected below. It is the linear approximation around
the local equilibrium.

Since, by detailed balance,
\begin{equation}
f^{(0)}({\vv})f^{(0)}({\vv}_{1})=f^{(0)}({\vv'})f^{(0)}({\vv'_1})
\label{Tri3.5}\end{equation}

\noindent for all such (${\vv}$, ${\vv_1}$), (${\vv'}$, ${\vv'_1}$)
which are related to each other by conservation laws, we have,
\begin{equation}
M_{Q}[f^{(0)},\mu ]=0, \: \hbox{ for any }\mu .
\end{equation}

\noindent Further, by virtue of conservation laws,
\begin{equation}
M_{Q}[f,\hat{P}^{(0)}\mu ]=0, \: \mbox{ for any }  f. \label{Tri3.7}
\end{equation}

\noindent From (\ref{Tri3.5})-(\ref{Tri3.7}) it follows,
\begin{eqnarray}
&&M_{Q}[f^{(0)}(1+\varphi ),\mu ]=M_{Q}[\varphi
,(\hat{1}-\hat{P}^{(0)})\mu ]\\ && = -\int w({\vv'},{\vv'_1} \mid
{\vv},{\vv_1}) f^{(0)}({\vv})f^{(0)}({\vv_1})
\left\{(1-\hat{P}^{(0)})\mu ({\vv})\right\}  \,\D {\vv'} \,\D
{\vv'_1} \,\D {\vv_1} \,\D {\vv}. \nonumber \label{Tri3.8}
\end{eqnarray}

\noindent We used notation,
\begin{equation}
\left\{ \psi ({\vv})\right\} =\psi ({\vv})+\psi ({\vv}_{1})-\psi
({\vv'})-\psi ({\vv'_1}). \label{Tri3.9}\end{equation}

\noindent Also,  it proves  convenient  to  introduce the
microscopic  density  of the scattering rate, $\mu _{Q}({\vv})$:
\begin{equation}
\mu _{Q}({\vv})=\int w({\vv'},{\vv'_1} \mid
{\vv},{\vv}_{1})f^{(0)}({\vv}_{1})\left\{(1-\hat{P}^{(0)})\mu
({\vv})\right\} \,\D {\vv'} \,\D {\vv'_1} \,\D {\vv}_{1}.
\label{Tri3.10}\end{equation}

\noindent Then,
\begin{equation}
M_{Q}[\varphi ,\mu ]=-(\varphi ,\mu _{Q}),
\label{Tri3.11}\end{equation}

\noindent where $(\cdot ,\cdot )$ is the $L_2$ scalar product with
the weight $f^{(0)}$ (\ref{Tri2.18}). This is a natural scalar
product in the space of functions $\varphi$ (\ref{Tri3.4})
(multipliers), and it is obviously related to the entropic scalar
product in the space of distribution functions at the local
equilibrium $f^{(0)}$, which is the $L_2$ scalar product with the
weight $(f^{(0)})^{-1}$.

Now, we obtain  transport  equations  for  the scattering rates
(\ref{Tri3.11}). We write down the time derivative of the collision
integral due to the Boltzmann equation,
\begin{equation}
\partial _{t}Q(f,f)({\vv}) =\hbox{ {\it \^{T}}Q}(f,f)({\vv}) +\hbox{ {\it \^{R}}Q}(f,f)({\vv}),
\label{Tri3.12}\end{equation}

\noindent where
\begin{eqnarray}
\hat{T} Q (f,f)({\vv})&& =\int w({\vv'},{\vv'_1} \mid
{\vv},{\vv_{1}}) \left[ f({\vv})v_{1k}\partial _{k}f({\vv}_{1}) +
f({\vv}_{1})v_{k}\partial _{k}f({\vv}) \right. \nonumber
\\  - && \left.  f({\vv'}) v'_{1k}\partial _{k}f({\vv'_1}) - f({\vv'_1}) v'_k
\partial_{k} f({\vv'}) \right]   \,\D {\vv'} \,\D {\vv'_1} \,\D {\vv_1} \,\D {\vv};
\label{Tri3.13} \\ \hat{R} Q (f,f)({\vv})&& = \int w
({\vv'},{\vv'_1} \mid {\vv},{\vv_1}) \left[
Q(f,f)({\vv'})f({\vv'_1}) + Q(f,f)({\vv'_1})f({\vv'}) \right.
\nonumber \\  -&& \left. Q(f,f)({\vv}_{1})f({\vv}) -
Q(f,f)({\vv})f({\vv}_{1}) \right] \,\D {\vv'} \,\D {\vv'_1} \,\D
{\vv_{1}} \,\D {\vv}. \label{Tri3.14}\end{eqnarray}

\noindent Using the representation,
\begin{eqnarray}
&&\partial _{k}f^{(0)}({\vv})=A_{k}({\vv})f^{(0)}({\vv});  \\ &&
A_{k}({\vv})=\partial _{k}\ln (nT^{-3/2})+{m\over k_{\rm
B}T}(v_{i}-u_{i})\partial _{k}u_{i}+{m({\vv}-{\uu})^{2}\over 2k_{\rm
B}T}\partial _{k}\ln T, \nonumber \label{Tri3.15}
\end{eqnarray}
\noindent and after some simple transformations using the relation
\begin{equation}
\left\{ A_{k}({\vv})\right\} =0, \label{Tri3.16}
\end{equation}
\noindent in linear with respect to $\varphi $ deviation from
$f^{(0)}$ (\ref{Tri3.4}), we obtain  in (\ref{Tri3.12}):
\begin{eqnarray}
\hat{T} Q (f,f)({\vv})&=&\partial_k  \int w({\vv'},{\vv'_1} \mid
{\vv},{\vv_1}) f^{(0)}({\vv_1}) f^{(0)}({\vv})\left\{ v_{k}\varphi
({\vv})\right\} \,\D {\vv'_1} \,\D {\vv'} \,\D {\vv_1} \nonumber
\\ && +\int w({\vv'},{\vv'_1} \mid {\vv},{\vv_1}) f^{(0)}({\vv_1}) f^{(0)}({\vv})\left\{
v_{k}A_{k}({\vv})\right\} \,\D {\vv'} \,\D {\vv'_1}
 \,\D {\vv_1} \nonumber
\\ && + \int w({\vv'},{\vv'_1} \mid {\vv},{\vv}_{1})f^{(0)}({\vv})f^{(0)}({\vv_1})
\left[ \varphi ({\vv})A_{k}({\vv}_{1})(v_{1k}-v_{k}) \right.
\nonumber \\ && + \varphi
({\vv_1})A_{k}({\vv})(v_{k}-v_{1k})+\varphi
({\vv'})A_{k}({\vv'_1})(v'_k- v'_{1k}) \nonumber \\ && + \left.
\varphi ({\vv'_1}) A_{k}({\vv'})(v'_{1k}-v'_{k})\right]  \,\D
{\vv'_1} \,\D {\vv'} \,\D {\vv_1}; \label{Tri3.17} \\ \hat{R}
Q(f,f)({\vv})& =& \int w({\vv'},{\vv'_1} \mid {\vv},{\vv_1})
f^{(0)}({\vv})f^{(0)}({\vv_1}) \left\{ \xi ({\vv})\right\} \,\D
{\vv'_1} \,\D {\vv'} \,\D {\vv_1}; \nonumber \\  \xi ({\vv}) & = &
\int w({\vv'},{\vv'_1} \mid {\vv},{\vv_1}) f^{(0)}({\vv_1}) \left\{
\varphi ({\vv})\right\} \,\D {\vv'_1}
 \,\D {\vv'} \,\D {\vv_1}; \label{Tri3.18}
\\
\partial_{t} Q(f,f)({\vv}) &  &  \\ &=&  -\partial_t \int w({\vv'},{\vv'_1} \mid
{\vv},{\vv_1}) f^{(0)}({\vv})f^{(0)}({\vv_1}) \left\{ \varphi
({\vv})\right\} \,\D {\vv'} \,\D {\vv'_1} \,\D {\vv_1}.
\label{Tri3.19}\nonumber
\end{eqnarray}

Let us use two identities:

\noindent 1. From the conservation laws it follows
\begin{equation} \left\{ \varphi ({\vv})\right\} =\left\{ (\hat{1}-\hat{P}^{(0)})\varphi
({\vv})\right\}. \label{Tri3.21}\end{equation} \noindent 2. The
symmetry property of the kernel $w$ (\ref{Tri3.2}) which follows
from (\ref{Tri3.2}), (\ref{Tri3.5})
\begin{eqnarray}
&&\int w({\vv'},{\vv'_1} \mid
{\vv},{\vv_1})f^{(0)}({\vv_1})f^{(0)}({\vv})g_{1}({\vv})\left\{
g_{2}({\vv})\right\} \,\D {\vv'} \,\D {\vv'_1} \,\D {\vv_1} \,\D
{\vv}
\\ &&= \int w({\vv'},{\vv'_1} \mid {\vv},{\vv_1}) f^{(0)}({\vv_1}) f^{(0)}({\vv})
g_{2}({\vv})\left\{ g_{1}({\vv})\right\} \,\D {\vv'} \,\D {\vv'_1}
\,\D {\vv_1} \,\D {\vv}. \label{Tri3.20} \nonumber
\end{eqnarray}
\noindent It is valid for any two functions $g_{1}$, $g_{2}$
ensuring existence  of the integrals, and also using the first
identity.

Now, multiplying (\ref{Tri3.17})-(\ref{Tri3.19}) by the microscopic
moment  density $\mu ({\vv})$, performing integration over ${\vv}$
(and  using identities (\ref{Tri3.21}), (\ref{Tri3.20})) we obtain
the required transport equation for the scattering rate  in the
linear neighborhood of the local equilibrium:
\begin{eqnarray}
&&-\partial _{t}\Delta M_{Q}[\varphi ,\mu ] \equiv  -\partial
_{t}(\varphi ,\mu _{Q}) \nonumber
\\ &&=(v_{k}A_{k}({\vv}),\mu _{Q}((\hat{1}-\hat{P}^{(0)})\mu ({\vv})))\nonumber
\\&& +  \partial _{k}(\varphi ({\vv})v_{k},\mu _{Q}((\hat{1}-\hat{P}^{(0)})\mu ({\vv})))+\int
w({\vv'},{\vv'_1}\mid {\vv},{\vv_1}) f^{(0)}({\vv_1})f^{(0)}({\vv})
\nonumber
\\ && \times  \left\{ (\hat{1}-\hat{P}^{(0)})\mu ({\vv})\right\} A_{k}({\vv_1})(v_{1k}-v_{k})\varphi
({\vv}) \,\D {\vv'} \,\D {\vv'_1} \,\D {\vv_1}d{\vv} \nonumber \\ &&
+ \left(\xi ({\vv}),\mu _{Q} \left((\hat{1}-\hat{P}^{(0)})\mu
({\vv}) \right)\right). \label{Tri3.22}
\end{eqnarray}

The chain of equations (\ref{Tri3.22}) for scattering rates is a
counterpart of the hydrodynamic moment chain (\ref{Tri2.31}).
Below we call (\ref{Tri3.22}) the {\it second chain}, and
(\ref{Tri2.31}) - the {\it first chain}.  Equations of the second
chain are coupled in  the same way as the first one: the last term
in the right part of (\ref{Tri3.22}) $(\xi ,\mu
_{Q}((\hat{1}-\hat{P}^{(0)})\mu ))$ depends on the whole totality
of moments and scattering rates and may  be treated as a new
variable. Therefore, generally speaking, we have an infinite
sequence of chains of increasingly higher orders. Only in the case
of a special choice of the collision model -- Maxwell potential
$U=-\kappa r^{-4} $ -- this sequence degenerates: the second and
the higher-order chains are equivalent to the first (see below).

Let  us  restrict  our consideration  to   the   first   and
second hydrodynamic chains. Then a deviation from the local
equilibrium  state and  transition  to a closed  macroscopic
description
 may be performed in three different ways for the  microscopic moment density $\mu ({\vv})$.
First, one can specify the moment $\hat{M}[\mu ]$ and perform a
closure of the chain (\ref{Tri2.31}) by the triangle method given
in previous subsections. This leads to Grad's moment method.
Second, one can specify scattering rate $\hat{M}_{Q}[\mu ]$ and
perform a closure of the second hydrodynamic chain
(\ref{Tri3.22}). Finally, one  can consider simultaneously both
$\hat{M}[\mu ]$ and $\hat{M}_{Q}[\mu ]$ (mixed chain).
Quasiequilibrium distribution functions corresponding to the last
two variants will be constructed in the  following  subsection.
The hard spheres model (H.S.) and Maxwell's molecules (M.M.) will
be considered.

\section{Distribution functions  of the second quasiequilibrium
approximation for scattering rates}

\subsection{First five moments and collision stress tensor}

Elsewhere  below  the local equilibrium  $f^{(0)} (\ref{Tri2.18})$
is chosen as the first quasi\-equilibrium approximation.
\par
Let us choose $\nu _{ik}=mv_{i}v_{k} (\ref{Tri2.20})$ as the
microscopic density $\mu ({\vv})$ of the second  quasiequilibrium
state.  Let  us  write   down   the corresponding scattering rate
(collision stress tensor) $\Delta _{ik}$  in the form,
\begin{equation}
\Delta _{ik}=-(\varphi ,\nu _{Qik}), \label{Tri4.1}\end{equation}
\noindent where
\begin{eqnarray}
&&\nu _{Qik} ({\vv})=m \int w({\vv'},{\vv'_1} \mid {\vv_1},{\vv})
f^{(0)}({\vv_1}) \nonumber \\ && \times \left\{
(v_{i}-u_{i})(v_{k}-u_{k}) - {1\over 3}\delta
_{ik}({\vv}-{\uu})^{2}\right\} \,\D {\vv'} \,\D {\vv'_1} \,\D
{\vv_1} \label{Tri4.2}\end{eqnarray}

\noindent is the microscopic density of the scattering rate $\Delta
_{ik}$.

The quasiequilibrium distribution function of the second
quasiequilibrium approximation  for fixed
 scattering rates (\ref{Tri4.1}) is determined as the  solution  to  the problem
\begin{eqnarray}
(\varphi ,\varphi )\rightarrow \min \hbox{  for} \nonumber\\
(\varphi ,\nu _{Qik})=-\Delta _{ik}. \label{Tri4.3} \end{eqnarray}

\noindent The method of Lagrange multipliers yields
\begin{eqnarray}
\varphi ({\vv})=\lambda _{ik}\nu _{Q ik}({\vv}), \nonumber \\
\lambda _{ik}(\nu _{Qik},\nu_{Qls})=\Delta _{ls},
\label{Tri4.4}\end{eqnarray}

\noindent where $\lambda_{ik}$ are the Lagrange multipliers.

In the examples of collision models considered below  (and  in
general, for centrally symmetric interactions) $ \nu _{Q ik}$ is of
the form
\begin{equation}
\nu _{Qik}({\vv})= (\hat{1}-\hat{P}^{(0)})\nu _{ik}({\vv})\Phi
(({\vv}-{\uu})^{2}), \label{Tri4.5}\end{equation} \noindent where
$(\hat{1}-\hat{P}^{(0)})\nu _{ik}$ is determined by relationship
(\ref{Tri2.21}) only, and function $\Phi $ depends only on the
absolute value of the peculiar velocity $({\vv}-{\uu})$. Then
\begin{eqnarray}
\lambda _{ik}&=&r\Delta _{ik};\nonumber \\ r^{-1}&=&(2/15)
\left(\Phi ^{2}(({\vv}-{\uu})^{2}),({\vv}-{\uu})^{4} \right),
\label{Tri4.6}\end{eqnarray}

\noindent and the distribution  function   of   the   second
quasiequilibrium approximation for scattering rates (\ref{Tri4.1})
is given by the expression
\begin{equation}
f=f^{(0)}(1+r\Delta _{ik}\mu_{Qik}). \label{Tri4.7}\end{equation}

The form of the function $\Phi (({\vv}-{\uu})^{2})$, and the value
of the parameter $r$  are determined by the model of particle's
interaction. In  the Appendix, they  are found for hard spheres
and  Maxwell molecules models (see
(\ref{TriA.14})-(\ref{TriA.19})). The distribution function
(\ref{Tri4.7}) is given by the following expressions:

\noindent For Maxwell molecules:
\begin{eqnarray}
&&f=f^{(0)} \nonumber \\ && \; \;  \times  \left\{
1+\mu^{\hbox{\scriptsize{M.M.}}}_{0}m(2P^{2}k_{\rm
B}T)^{-1}\Delta_{ik} \left( (v_{i}-u_{i})(v_{k}-u_{k}) - {1\over
3}\delta _{ik}({\vv}-{\uu})^{2} \right) \right\}, \nonumber
\\&& \mu^{\hbox{\scriptsize{M.M.}}}_{0}={k_{\rm B}T \sqrt{2m} \over 3\pi A_{2}(5)\sqrt{\kappa}} , \label{Tri4.8}
\end{eqnarray}

\noindent where $\mu^{\hbox{\scriptsize{M.M.}}}_{0}$  is viscosity
coefficient  in the first approximation of the Chapman--Enskog
method (it is exact in the case of Maxwell molecules), $\kappa $
is a force constant, $A_{2}(5)$ is a number, $A_{2}(5)\approx
0.436$ (see \cite{Chapman});

For the hard spheres model:
\begin{eqnarray}
&&f=f^{(0)}  \nonumber \\ && \; \; \times \left\{ 1+{2\sqrt{2}
\tilde{r} m \mu^{\hbox{\scriptsize{H.S.}}}_{0}\over 5P^{2}k_{\rm
B}T}\Delta _{ik} \int^{-1}_{+1} \exp
\left\{-{m({\vv}-{\uu})^{2}\over 2k_{\rm B}T}y^{2} \right\}
(1-y^{2})(1+y^{2}) \right. \nonumber \\ && \;\; \times \left. \left(
{m({\vv}-{\uu})^{2}\over 2k_{\rm B}T}(1-y^{2})+2 \right) \,\D y
\left((v_{i}-u_{i})(v_{k}-u_{k}) - {1\over 3}\delta
_{ik}({\vv}-{\uu})^{2} \right) \right\}, \nonumber \\
&&\mu^{\hbox{\scriptsize{H.S.}}}_{0}=(5 \sqrt{k_{\rm B} Tm})/(16
\sqrt{\pi} \sigma ^{2}), \label{Tri4.9}
\end{eqnarray}

\noindent where $\tilde{r}$ is a number represented as follows:
\begin{eqnarray}
&&\tilde{r}^{-1}={1\over 16} \int_{-1}^{+1} \int_{-1}^{+1} \alpha
^{-11/2} \beta (y)\beta (z)\gamma (y)\gamma (z)\nonumber \\ && \;\;
\times  (16\alpha ^{2}+28\alpha (\gamma (y)+\gamma (z))+63\gamma
(y)\gamma (z))  \,\D y \,\D z, \label{Tri4.10} \\ &&\alpha
=1+y^{2}+z^{2},\qquad \beta (y)=1+y^{2},\qquad \gamma
(y)=1-y^{2}.\nonumber
\end{eqnarray}

Numerical value of $\tilde{r}^{-1}$ is 5.212, to third decimal point
accuracy.

In the mixed description, the distribution  function   of   the
second  quasiequilibrium approximation under fixed values of the
moments  and of the scattering rates corresponding to the
microscopic density (\ref{Tri2.20}) is determined as a solution of
the problem
\begin{eqnarray}
&&(\varphi ,\varphi )\rightarrow \min \: \hbox{ for } \label{Tri4.11} \\
&&((\hat{1}-\hat{P}^{(0)})\nu _{ik},\varphi)=\sigma _{ik},
\nonumber \\ &&(\nu _{Q ik},\varphi )=\Delta _{ik}. \nonumber
\end{eqnarray}

 Taking into account the relation
(\ref{Tri4.5}),  we obtain the solution of the problem
(\ref{Tri4.11}) in the form,
\begin{equation}
\varphi ({\vv})=(\lambda _{ik}\Phi (({\vv}-{\uu})^{2})+\beta
_{ik})((v_{i}-u_{i})(v_{k}-u_{k}) - (1/3)\delta
_{ik}({\vv}-{\uu})^{2}). \label{Tri4.12}\end{equation}

Lagrange multipliers $\lambda _{ik}, \beta _{ik}$  are  determined
from  the system of linear equations,
\begin{eqnarray} {ms^{-1}\lambda _{ik}+2Pk_{\rm B}Tm^{-1}\beta _{ik}=\sigma _{ik},} \nonumber \\
{mr^{-1}\lambda_{ik}+ms^{-1}\beta _{ik}=\Delta _{ik},}
\label{Tri4.13}
\end{eqnarray}

\noindent where
\begin{equation}
s^{-1}=(2/15)(\Phi (({\vv}-{\uu})^{2}),({\vv}-{\uu})^{4}).
\label{Tri4.14}\end{equation}

If the solvability condition of the system (\ref{Tri4.13}) is
satisfied,
\begin{equation}
D=m^{2}s^{-2}-2Pk_{\rm B}Tr^{-1}\neq 0,
\label{Tri4.15}\end{equation}

\noindent then the distribution  function  of   the   second
quasiequilibrium approximation exists and takes the form
\begin{eqnarray}
f&=&f^{(0)}\left\{ 1+(m^{2} s^{-2}-2Pk_{\rm B}Tr^{-1})^{-1} \right.
\\ && \times [(ms^{-1}\sigma _{ik}-2Pk_{\rm B}Tm^{-1}\Delta
_{ik})\Phi (({\vv}-{\uu})^{2})\nonumber \\ && + \left.
(ms^{-1}\Delta _{ik}-mr^{-1}\sigma
_{ik})]((v_{i}-u_{i})(v_{k}-u_{k}) - (1/3)\delta
_{ik}({\vv}-{\uu})^{2})\right\} .  \nonumber \label{Tri4.16}
\end{eqnarray}

The condition (\ref{Tri4.15}) means independence of the set  of
moments $\sigma _{ik}$ from the scattering rates $\Delta _{ik}$. If
this condition is not satisfied,  then the scattering rates $\Delta
_{ik}$ can be represented in the form of linear combinations of
$\sigma _{ik}$ (with coefficients depending on the hydrodynamic
moments). Then the closed by means of (\ref{Tri4.7}) equations of
the second chain are equivalent to the ten moment Grad equations,
while the mixed chain  does not exist. This happens only in the case
of  Maxwell molecules. Indeed, in this case
$$ s^{-1}=2P^{2}k_{\rm B}T(m^{2}\mu
^{\hbox{\scriptsize{M.M.}}}_{0})^{-1}; D=0. $$ \noindent The
transformation changing $\Delta _{ik}$ to $\sigma _{ik}$ is

\begin{equation}
\mu ^{\hbox{\scriptsize{M.M.}}}_{0}\Delta _{ik}P^{-1}=\sigma _{ik}.
\label{Tri4.17}\end{equation}

\noindent For hard spheres:
\begin{equation}
s^{-1}={5P^{2}k_{\rm B}T\over 4\sqrt{2}
\mu^{\hbox{\scriptsize{H.S.}}}_{0} m^{2}}\cdot \tilde{s}^{-1}, \;
\tilde{s}^{-1}=\int_{-1}^{+1} \gamma (y)(\beta (y))^{-7/2} \left(
\beta (y)+{7\over 4}\gamma (y) \right)  \,\D y.
\label{Tri4.18}\end{equation}

The numerical value of $\tilde{s}^{-1}$ is 1.115 to third decimal
point. The condition (\ref{Tri4.14}) takes the form,
\begin{equation}
D={25\over 32}\left({P^{2}k_{\rm B}T\over m\mu
^{\hbox{\scriptsize{H.S.}}}_{0}}
\right)^{2}(\tilde{s}^{-2}-\tilde{r}^{-1})\neq 0.
\label{Tri4.19}\end{equation}

\noindent Consequently, for  the  hard spheres  model  the
distribution function of the  second quasiequilibrium approximation
of  the mixed chain exists and is determined by the expression
\begin{eqnarray}
f&=&f^{(0)}\left\{ 1+m(4Pk_{\rm
B}T(\tilde{s}^{-2}-\tilde{r}^{-1}))^{-1}  \right. \nonumber \\&&
\times \left[ \left( \sigma_{ik} \tilde{s}^{-1} - {8\sqrt{2}\over
5P}\mu ^{\hbox{\scriptsize{H.S.}}}_{0}\Delta_{ik} \right)
\int_{-1}^{+1} \exp \left( -{m({\vv}-{\uu})^{2}\over 2k_{\rm B}
T}y^{2} \right)  \right. \nonumber \\ && \times (1-y^{2})(1+y^{2})
\left( {m({\vv}-{\uu})^{2}\over 2k_{\rm B}T} (1-y^{2})+2 \right)
\,\D y\nonumber \\&& \left. \left.+ 2 \left( \tilde{s}^{-1}\cdot {8
\sqrt{2} \over 5P} \mu^{\hbox{\scriptsize{H.S.}}}_0 \Delta_{ik} -
\tilde{r}^{-1} \sigma_{ik} \right) \right] \right.   \nonumber \\
&&\left. \left. \times ( (v_i -u_i)(v_k-u_k) - {1 \over 3}
\delta_{ik} ({\vv}-{\uu})^2 \right) \right\}. \label{Tri4.20}
\end{eqnarray}

\subsection{First five moments, collision stress tensor, and
collision heat  flux  vector}

Distribution  function   of   the   second  quasiequilibrium
approximation which takes into account the collision heat  flux
vector $Q$ is constructed in a similar way. The microscopic density
$\xi _{Q i}$ is
\begin{equation}
\xi _{Q i}({\vv})=\int w({\vv'},{\vv'_1}\mid
{\vv},{\vv_1})f^{(0)}({\vv_1})\left\{(\hat{1}-\hat{P}^{(0)})
{v^{2}_{i}v\over 2}\right\} \,\D {\vv'} \,\D {\vv'_1} \,\D {\vv_1}.
\label{Tri4.21}\end{equation}

The desired distribution functions are the solutions  to  the
following optimization problems:  for the second chain it is the
solution to the  problem  (\ref{Tri4.3})  with the additional
constraints,
\begin{equation}
m(\varphi ,\xi_{Qi})=Q_{i}. \label{Tri4.22}
\end{equation}

\noindent For the  mixed  chain, the distribution functions  is the
solution to  the  problem (\ref{Tri4.11})  with additional
conditions,
\begin{eqnarray}
m(\varphi ,\xi _{Qi})&=&Q_{i}, \\ m(\varphi
,(\hat{1}-\hat{P}^{(0)})\xi _{i})&=&q_{i}. \label{Tri4.23}
\end{eqnarray}

Here  $\xi _{i}=v_{i}v^{2}/2$ (see (\ref{Tri2.23})). In the Appendix
functions $\xi _{Q i}$ are found for Maxwell  molecules and hard
spheres (see (\ref{TriA.19})-(\ref{TriA.24})). Since
\begin{eqnarray}&& (\xi _{Qi},\nu _{Q kj})=((\hat{1}-\hat{P}^{(0)})\xi _{i},\nu
_{Qkj}) \nonumber \\ && = (\xi _{Q i},(\hat{1}-\hat{P}^{(0)})\nu
_{kj})=((\hat{1}-\hat{P}^{(0)})\xi _{i},(\hat{1}-\hat{P}^{(0)})\nu
_{kj})=0,
\end{eqnarray}

\noindent the conditions (\ref{Tri4.22}) are linearly independent
from  the  constraints  of  the problem (\ref{Tri4.3}), and  the
conditions  (\ref{Tri4.23})  do  not  depend  on  the constraints of
the problem (\ref{Tri4.11}).

Distribution  function   of   the   second  quasiequilibrium
approximation of the second chain for fixed $\Delta _{ik}, Q_{i}$ is
of the form,
\begin{equation}
f=f^{(0)}(1+r\Delta _{ik}\nu _{Qik}+\eta Q_{i}\xi _{Q i}).
\label{Tri4.24}\end{equation}

\noindent The parameter $\eta $ is determined by the relation
\begin{equation}
\eta ^{-1}=(1/3)(\xi _{Qi},\xi _{Qi}). \label{Tri4.25}\end{equation}

According to (\ref{TriA.23}), for Maxwell molecules
\begin{equation}
\eta ={9m^{3}(\mu ^{\hbox{\scriptsize{M.M.}}}_{0})^{2}\over
10P^{3}(k_{\rm B}T)^{2}}, \label{Tri4.26}\end{equation}

\noindent and the distribution function (\ref{Tri4.24}) is
\begin{eqnarray}
&&f=f^{(0)}  \nonumber \\ && \; \times   \left\{ 1+\mu
^{\hbox{\scriptsize{M.M.}}}_{0}m(2P^{2}k_{\rm B}T)^{-1}\Delta
_{ik}((v_{i}-u_{i})(v_{k}-u_{k}) - (1/3)\delta
_{ik}({\vv}-{\uu})^{2}) \right. \nonumber \\ && \; + \left. \mu
^{\hbox{\scriptsize{M.M.}}}_{0} m(P^{2}k_{\rm B}T)^{-1}(v_{i}-u_{i})
\left({m({\vv}-{\uu})^{2}\over 5k_{\rm B}T} - 1 \right) \right\}.
\label{Tri4.27}
\end{eqnarray}

\noindent For hard spheres (see Appendix)
\begin{equation}
\eta =\tilde{\eta } {64m^{3}(\mu
^{\hbox{\scriptsize{H.S.}}}_{0})^{2}\over 125P^{3}(k_{\rm B}T)^{2}},
\label{Tri4.28}
\end{equation}

\noindent where $\eta $ is a number equal to 16.077 to third decimal
point accuracy.

The distribution function (\ref{Tri4.24}) for hard spheres  takes
the form
\begin{eqnarray}
f&=&f^{(0)}\left\{ 1+{2\sqrt{2} \tilde{r} m
\mu^{\hbox{\scriptsize{H.S.}}}_{0}\over 5P^{2}k_{\rm B}T}
\Delta_{ik} \int_{-1}^{+1} \exp \left(-{m({\vv}-{\uu})^{2}\over
2k_{\rm B}T}y^{2} \right)  \beta (y)\gamma (y) \right. \nonumber
\\&& \times \left( {m({\vv}-{\uu})^{2}\over 2k_{\rm B}T}\gamma (y)+2
\right)  \,\D y \left( (v_{i}-u_{i})(v_{k}-u_{k}) - {1\over 3}\delta
_{ik}({\vv}-{\uu})^{2} \right) \nonumber \\ && + {2
\sqrt{2}\tilde{\eta} m^{3} \mu^{\hbox{\scriptsize{H.S.}}}_{0} \over
25P^{2}(k_{\rm B}T)^{2}} Q_i \left[ (v_{i}-u_{i}) \left( (
{\vv}-{\uu})^{2} - {5k_{\rm B}T\over m} \right) \right. \nonumber \\
&&  \times  \int_{-1}^{+1} \exp \left(-{m({\vv}-{\uu})^{2}\over
2k_{\rm B}T}y^{2}\right) \beta (y)\gamma (y)
\left({m({\vv}-{\uu})^{2}\over 2k_{\rm B}T}\gamma (y)+2 \right) \,\D
y \nonumber \\&& + (v_{i}-u_{i})({\vv}-{\uu})^{2} \int^{+1}_{-1}
\exp \left(-{m({\vv}-{\uu})^{2}\over 2k_{\rm B}T}y^{2} \right) \beta
(y)\gamma (y)\nonumber \\ && \left. \left. \times
 \left( \sigma (y){m({\vv}-{\uu})^{2}\over 2k_{\rm B}T}+\delta (y) \right)   \,\D y  \right]
\right\}. \label{Tri4.29}
\end{eqnarray}
\noindent The functions $\beta (y), \gamma (y), \sigma (y)$ and
$\delta (y)$ are
\begin{equation}
\beta (y)=1+y^{2}, \: \gamma (y)=1-y^{2}, \: \sigma
(y)=y^{2}(1-y^{2}), \: \delta (y)=3y^{2}-1. \label{Tri4.30}
\end{equation}
\par
The condition of existence  of  the  second quasiequilibrium
approximation of the mixed chain (\ref{Tri4.15}) should be
supplemented  with the requirement
\begin{equation}
R=m^{2}\tau^{-2} - {5P(k_{\rm B}T)^{2}\over 2m}\eta ^{-1}\neq 0.
\label{Tri4.31}\end{equation} \noindent Here
\begin{equation}
\tau^{-1}={1\over 3} \left( (\hat{1}-\hat{P}^{(0)}){v^{2}_{i}v\over
2},\xi _{Qi}({\vv}) \right). \label{Tri4.32}\end{equation} \noindent
For Maxwell molecules $$ \tau^{-1}=(5P^{2}k^{2}_{B}T^{2})/(3\mu
^{\hbox{\scriptsize{M.M.}}}_{0}m^{3}), $$ \noindent and the
solvability condition (\ref{Tri4.31}) is not satisfied. Distribution
function of the second quasiequilibrium approximation of mixed chain
does not exist for Maxwell molecules. The variables $Q_{i}$ are
changed to $q_{i}$ by the transformation
\begin{equation}
3\mu ^{\hbox{\scriptsize{M.M.}}}_{0}Q_{i}=2Pq_{i}.
\label{Tri4.33}\end{equation} For hard spheres,
\begin{equation}
\tau^{-1}=\tilde{\tau}^{-1}={25(Pk_{\rm B}T)^{2}\over 8 \sqrt{2}
m^{3}\mu ^{\hbox{\scriptsize{H.S.}}}_{0}},
\label{Tri4.34}\end{equation} \noindent where
\begin{eqnarray}
\tilde{\tau}^{-1}&=&{1\over 8} \int^{+1}_{-1}\beta ^{-9/2}(y)\gamma
(y) \{ 63(\gamma (y)+\sigma (y)) \nonumber \\ &&+7\beta
(y)(4-10\gamma (y) + 2\delta (y)-5\sigma (y)) \nonumber \\ && +
\beta ^{2}(y)(25\gamma (y)-10\delta (y)-40)+20\beta ^{3}(y) \} \,\D
y. \label{Tri4.35}
\end{eqnarray}
\noindent The numerical value of $\tilde{\tau}^{-1}$ is about 4.322.
Then the condition (\ref{Tri4.31}) is verified: $$ R\approx
66m^{-4}(Pk_{\rm B}T)^{4}(\mu ^{\hbox{\scriptsize{H.S.}}}_{0})^{2}.
$$

\noindent Finally,  for the fixed  values  of $\sigma _{ik}, \Delta
_{ik}, q_{i}$  and $Q_{i}$   the distribution   function of the
second    quasiequilibrium approximation of the second chain for
hard spheres is of the form,
\begin{eqnarray}
&& f=f^{(0)} \left\{ 1+{m\over 4Pk_{\rm
B}T}(\tilde{s}^{-2}-\tilde{r}^{-1}) ^{-1} \right.
 \nonumber \\ && \; \times  \left[ \left( \tilde{s}^{-1}\sigma _{ik} -
{8\sqrt{2}\over 5P} \mu^{\hbox{\scriptsize{H.S.}}}_{0} \Delta_{ik}
\right) \int^{+1}_{-1}\exp \left(-{m({\vv}-{\uu})^{2}\over 2k_{\rm
B}T}y^{2} \right)  \right. \nonumber
\\  &&  \;\left.  \times \beta (y)\gamma (y) \left({m({\vv}-{\uu})^{2}\over 2k_{\rm B}T} \gamma (y)+2 \right)
 \,\D y+2 \left( \tilde{s}^{-1}{8\sqrt{2}\over 5P} \mu^{\hbox{\scriptsize{H.S.}}}_{0} \Delta _{ik} -
\tilde{r}^{-1}\sigma_{ik} \right) \right]  \nonumber \\  && \;
\times  \left( (v_{i}-u_{i})(v_{k}-u_{k}) - {1\over 3}\delta
_{ik}({\vv}-{\uu})^{2} \right) \nonumber \\ && \; + {m^{2}\over
10(Pk_{\rm B}T)^{2}}(\tilde{\tau}^{-2}-\tilde{\eta }^{-1})^{-1}
\left[ \left( \tilde{\tau}^{-1}q_{i} - {4\sqrt{2} \over 5P}\mu
^{\hbox{\scriptsize{H.S.}}}_{0}Q_i \right) \right. \nonumber \\ &&
\; \times \left( (v_{i}-u_{i}) \left( ({\vv}-{\uu})^{2} - {5k_{\rm
B}T\over m} \right) \int^{+1}_{-1}
\exp\left(-{m({\vv}-{\uu})^{2}\over 2k_{\rm B}T}y^{2} \right)
\right. \nonumber
\\ && \; \times \beta (y)\gamma (y)\left({m ({\vv}-{\uu})^{2}\over 2k_{\rm B}T}\gamma (y)+ 2
\right) \,\D y+(v_{i}-u_{i})({\vv}-{\uu})^{2} \nonumber \\ && \;
\times   \left.\int^{+1}_{-1} \exp \left(-{m({\vv}-{\uu})^{2}\over
2k_{\rm B}T} y^{2}\right)
 \beta (y)\gamma (y) \left( {m({\vv}-{\uu})^{2}\over 2k_{\rm B}T}\sigma (y)+\delta (y)
\right)  \,\D y \right)
 \nonumber \\  && \; +\left. \left.
2 \left( {4\sqrt{2} \over 5P}
\mu^{\hbox{\scriptsize{H.S.}}}_{0}\tilde{\tau}^{-1}Q_{i} -
\tilde{\eta }^{-1}q_i \right) (v_{i}-u_{i}) \left(({\vv}-{\uu})^{2}
-  {5k_{\rm B}T\over m} \right) \right] \right\}. \label{Tri4.36}
\end{eqnarray}
Thus, the expressions (\ref{Tri4.8}), (\ref{Tri4.9}),
(\ref{Tri4.20}),  (\ref{Tri4.27}), (\ref{Tri4.29}) and
(\ref{Tri4.36}) give  distribution  functions  of  the   second
quasiequilibrium approximation of the second  and  mixed
hydrodynamic  chains  for Maxwell molecules and hard spheres. They
are analogues of ten- and thirteen-moment Grad approximations
(\ref{Tri2.22}), (\ref{Tri2.26}).
\par
The next step is to close the second and  mixed  hydrodynamic chains
by means of the found distribution functions.

\section{Closure of the second and mixed hydrodynamic chains}

\subsection{Second chain, Maxwell molecules} The distribution
function  of  the  second quasiequilibrium approximation under
fixed $\Delta _{ik}$  for  Maxwell  molecules (\ref{Tri4.8})
presents the simplest example of the closure of the first
(\ref{Tri2.31})  and second (\ref{Tri3.22}) hydrodynamic chains.
With the help of it, we  obtain from (\ref{Tri2.31}) the following
transport equations for the  moments  of the first (local
equilibrium) approximation:
\begin{eqnarray}
&&\partial_t \rho +\partial_i (u_i \rho)=0; \nonumber \\ &&\rho
(\partial_t u_k + u_i \partial_i u_k) +
\partial_k P + \partial_i ( P^{-1} \mu_0^{\hbox{\scriptsize{M.M.}}}  \Delta_{ik} ) =0;  \nonumber
\\ &&{3\over 2} (\partial _t P+ u _i
\partial_i P ) + {5 \over 2}P \partial_i u_i + P^{-1} \mu_0^{\hbox{\scriptsize{M.M.}}} \Delta_{ik} \partial_i u_k
=0. \label{Tri5.1}
\end{eqnarray}
\noindent Now, let us from the scattering rate transport chain
(\ref{Tri3.22})  find an equation for $\Delta _{ik}$ which  closes
the  system  (\ref{Tri4.1}).  Substituting (\ref{Tri4.8}) into
(\ref{Tri3.22}), we obtain after some computation:
\begin{eqnarray}
&&\partial_{t}\Delta _{ik}+\partial _{s}(u_{s}\Delta _{ik})+\Delta
_{is}
\partial_{s} u_{k}+\Delta _{ks}\partial _{s}u_{i}-{2\over 3}\delta _{ik}\Delta _{ls}\partial _{s}u_{l}
 \nonumber \\&& + P^{2}(\mu ^{\hbox{\scriptsize{M.M.}}}_{0})^{-1} \left(\partial _{i}u_{k}+\partial
_{k}u_{i}-{2\over 3}\delta _{ik}\partial _{s}u_s \right)\nonumber
\\&& +P(\mu ^{\hbox{\scriptsize{M.M.}}}_{0})^{-1}\Delta _{ik}+\Delta
_{ik}\partial _{s}u_{s}= 0. \label{Tri5.2}
\end{eqnarray}
\noindent For comparison, let us give  ten-moment  Grad  equations
obtained when closing  the  chain (\ref{Tri2.30})  by  the
distribution  functions (\ref{Tri2.22}):
\begin{eqnarray}
&&\partial_t \rho +\partial_i (u_i \rho)=0; \nonumber \\ &&\rho
(\partial_t u_k + u_i \partial_i u_k) +
\partial_k P + \partial_i \sigma_{ik}=0; \nonumber \\ &&{3\over 2} (\partial _t P+ u _i \partial_i P
) + {5 \over 2}P \partial_i u_i +\sigma_{ik}
\partial_i u_k =0; \label{Tri5.3} \\
&&\partial _{t}\sigma _{ik}+\partial _{s}(u_{s}\sigma _{ik})+P
\left(\partial _{i}u_{k}+\partial_{k} u_{i}-{2\over 3}\delta
_{ik}\partial_{s}u_s \right) \nonumber \\ && + \sigma _{is}\partial
_{s}u_{k}+\sigma _{ks}\partial _{s}u_{i}-{2\over 3}\delta
_{ik}\sigma _{ls}\partial _{s}u_{l}+P(\mu
^{\hbox{\scriptsize{M.M.}}}_{0})^{-1}\sigma _{ik}=0.
\end{eqnarray}
\noindent Using the explicit form of $\mu
^{\hbox{\scriptsize{M.M.}}}_{0} (\ref{Tri4.8})$, it is easy to
verify that the transformation (\ref{Tri4.17}) maps the systems
(\ref{Tri5.1}),
 (\ref{Tri5.2})  and
(\ref{Tri5.3}) into one another. This is a consequence of the
degeneration of the mixed hydrodynamic chain which  was already
discussed.  The systems (\ref{Tri5.1}), (\ref{Tri5.2}) and
(\ref{Tri5.3}) are essentially equivalent.  These specific
properties of Maxwell molecules result from the fact that for them
the microscopic densities $(\hat{1}-\hat{P}^{(0)})v_{i}v_{k}$ and
$(\hat{1}-\hat{P}^{(0)})v_{i}v^{2}$ are eigen functions of the
linearized collision integral.
\par

\subsection{Second chain, hard spheres}

We now turn our attention to  the  closure  of  the second and of
the mixed hydrodynamic chains for the hard spheres  model.
Substituting the  distribution  function  (\ref{Tri4.9})  into
(\ref{Tri2.30}) and (\ref{Tri3.22}), we obtain an analogue of the
systems (\ref{Tri5.1}) and (\ref{Tri5.2}) (second chain, hard
spheres):
\begin{eqnarray}
&&\partial_{t}\rho +\partial _{i}(u_{i}\rho )=0;  \\ && \rho
(\partial_{t}u_{k}+u_{i}\partial _{i}u_{k})+\partial
_{k}P+\tilde{r}\tilde{s}^{-1}\cdot {8\sqrt{2}\over 5}\partial
_{i}(\mu ^{\hbox{\scriptsize{H.S.}}}_{0}P^{-1}\Delta _{ik})=0;
\nonumber \\ && {3\over 2}(\partial _{t}P+u_{i}\partial
_{i}P)+{5\over 2}P\partial _{i}u_{i}+\tilde{r}\tilde{s}^{-1}\cdot
{8\sqrt{2}\over 5}\mu ^{\hbox{\scriptsize{H.S.}}}_{0}P^{-1}\Delta
_{ik}\partial _{i}u_{k}=0; \nonumber \\ && \partial _{t}\Delta
_{ik}+\partial _{s}(u_{s}\Delta
_{ik})+\tilde{r}\tilde{a}_{1}(\partial _{s}u_{s})\Delta
_{ik}+{5\tilde{s}^{-1}P^{2}\over 8\sqrt{2} \mu
^{\hbox{\scriptsize{H.S.}}}_{0}} \left( \partial _{i}u_{k}+\partial
_{k}u_{i}-{2\over 3}\delta _{ik}\partial _{s}u_s \right) \nonumber
\\ && + \tilde{r}(\tilde{a}_{1}+\tilde{a}_{2})\left( \Delta
_{is}\partial _{s}u_{k}+\Delta _{ks}\partial _{s}u_{i}-{2\over
3}\delta _{ik}\Delta _{ls}\partial _{s}u_l \right) \nonumber \\ && +
\tilde{r}(\tilde{a}_{1}+\tilde{a}_{3}) \left( \Delta _{is}\partial
_{k}u_{s}+\Delta _{ks}\partial _{i}u_{s}-{2\over 3}\delta
_{ik}\Delta _{ls}\partial _{s}u_l \right)+ (
P\hat{r}\tilde{a}_{0}/\mu ^{\hbox{\scriptsize{H.S.}}}_{0})\Delta
_{ik}=0. \nonumber \label{Tri5.4}
\end{eqnarray}
\noindent The dimensionless parameters $\tilde{a}_{0},
\tilde{a}_{1}, \tilde{a}_{2}$ and $\tilde{a}_{3}$ are  determined
by the quadratures
\begin{eqnarray}
\tilde{a}_{1}&=&{1\over 16} \int_{-1}^{+1}\int_{-1}^{+1} \beta(y)
\beta(z) \gamma ^{2}(z)\gamma (y)\alpha ^{-13/2}(y,z) \nonumber \\
&& \times \{ 99\gamma (y)\gamma (z)(\gamma (z)-1)+18\alpha
(y,z)(2\gamma (z)(\gamma (z)-1) \nonumber \\ && + 4\gamma
(y)(4\gamma (z)-3))+8\alpha ^{2}(y,z)(4\gamma (z)-3) \}   \,\D y
\,\D z; \nonumber \\ \tilde{a}_{2}&=&{1\over 16}
\int_{-1}^{+1}\int_{-1}^{+1} \beta (y)\beta (z)\gamma (y)\gamma
^{2}(z)\alpha ^{-11/2}(y,z)\{ 63\gamma (y)\gamma (z)\nonumber
\\ && + 14\alpha (y,z)(3\gamma (y)+2\gamma (z))+24\alpha ^{2}(y,z)
\} \,\D y \,\D z; \nonumber
\\ \tilde{a}_{3}&=&{1\over 16} \int_{-1}^{+1}\int_{-1}^{+1} \alpha
^{-11/2}(y,z)\beta (y)\beta (z)\gamma (y)\gamma (z) \nonumber \\ &&
\times \{ 63\gamma(y)\gamma (z)(\gamma (z)-1)+14(2\gamma (z)(\gamma
(z)-1)\nonumber
\\ && +  \gamma (y)(3\gamma (z)-2))\alpha (y,z)+8\alpha ^{2}(y,z)(3\gamma (z)-2) \} dydz; \label{Tri5.5} \\
\tilde{a}_{0} & \approx & {1\over 1536 \sqrt{2}}
\int_{-1}^{+1}\int_{-1}^{+1}\int_{-1}^{+1} (\psi
(x,y,z))^{-13/2}\beta (x)\beta (y)\beta (z) \nonumber \\ && \times
\gamma (x)\gamma (y)\gamma (z)\{ 10395\gamma (x)\gamma (y)\gamma
(z)+3780\psi (x,y,z) \nonumber \\ && \times (\gamma (x)\gamma
(y)+\gamma (x)\gamma (z)+\gamma (y)\gamma (z))+1680\psi^{2}(x,y,z)
\nonumber \\ && \times (\gamma (x)+\gamma (y)+\gamma (z))+960
\psi^{3}(x,y,z) \}  \,\D x \,\D y \,\D z; \nonumber \\ && \psi
(x,y,z)=1+x^{2}+y^{2}+z^{2}.
\end{eqnarray}
\noindent Their numerical values are $\tilde{a}_{1}\approx 0.36$,
$\tilde{a}_{2}\approx 5.59$, $\tilde{a}_{3}\approx 0.38$,
$\tilde{a}_{0}\approx 2.92$ to second decimal point.
\par

\subsection{Mixed chain}

The closure of the mixed  hydrodynamic  chain  with  the functions
(\ref{Tri4.20}) gives the following modification of the system of
equations (\ref{Tri5.4}):
\begin{eqnarray}
&&\partial_{t}\rho +\partial_{i}(u_{i}\rho )=0; \nonumber \\ &&\rho
(\partial _{t}u_{k}+u_{i}\partial _{i}u_{k})+\partial _{k}P+\partial
_{i}\sigma _{ik}=0; \nonumber \\ &&{3\over 2}(\partial
_{t}P+u_{i}\partial _{i}P)+{5\over 2}P\partial _{i}u_{i}+\sigma
_{ik}\partial _{i}u_{k}=0; \nonumber
\\ &&
\partial_{t}\sigma _{ik}+\partial _{s}(u_{s}\sigma _{ik})+P \left( \partial _{i}u_{k}+\partial _{k}u_{i}
-{2\over 3}\delta _{ik}\partial _{s}u_s \right) \nonumber \\ && +
\sigma_{is}\partial _{s}u_{k}+\sigma _{ks}\partial _{s}u_{i}-{2\over
3}\delta _{ik}\sigma _{ls}\partial _{s}u_{l}+\Delta _{ik}=0;
\nonumber \\ &&\partial_{t}\Delta _{ik}+\partial _{s}(u_{s}\Delta
_{ik})+{5P^{2}\over \tilde{s}8\sqrt{2} \mu
^{\hbox{\scriptsize{H.S.}}}_{0}} \left(
\partial _{i}u_{k}+\partial _{k}u_{i}-{2\over 3}\delta _{ik}\partial _{s}u_s \right)\nonumber \\ && +
{5P\over 4\sqrt{2} \mu^{\hbox{\scriptsize{H.S.}}}_{0}
(\tilde{s}^{-2}-\tilde{r}^{-1})} \left\{ {\tilde{a}_{1}\over
2}(\partial _{s}u_{s})\alpha _{ik}\nonumber \right. \\  && +{1\over
2}(\tilde{a}_{1}+\tilde{a}_{2}) \left( \alpha _{is}\partial
_{s}u_{k}   + \alpha_{ks}
\partial_{s} u_{i} -{2\over 3}\delta_{ik} \alpha_{ls} \partial_{s} u_l \right)\nonumber \\ && +{1\over
2}(\tilde{a}_{1}+\tilde{a}_{3}) \left( \alpha _{is}\partial_{k}
u_{s}+ \alpha_{ks}
\partial_{i} u_{s}-{2\over 3} \delta_{ik}\alpha_{ls} \partial_{s} u_l \right)\nonumber \\ && +
\tilde{b}_{1} ( \partial_{s} u_{s}) \beta_{ik} +
(\tilde{b}_{1}+\tilde{b}_{2}) \left( \beta _{is}\partial
_{s}u_{k}+\beta _{ks} \partial_{s} u_{i}-{2\over 3}
\delta_{ik}\beta_{ls} \partial_{s} u_l \right)\nonumber \\ &&\left.
+  (\tilde{b}_{1}+\tilde{b}_{3}) \left( \beta _{is}\partial
_{k}u_{s}+\beta _{ks}\partial _{i}u_{s}-{2\over 3}\delta _{ik} \beta
_{ls} \partial _{s} u_l \right) \right\}   \nonumber \\ && +
{5P^{2}\over 8\sqrt{2} (\mu
^{\hbox{\scriptsize{H.S.}}}_{0})^{2}(\tilde{s}^{-2}-\tilde{r}^{-1})}
\left\{ {5\over 8\sqrt{2}
\tilde{r} }\beta_{ik}+\tilde{a}_{0} \alpha_{ik} \right\}=0; \label{Tri5.6} \\
&&\alpha_{ik}=\tilde{s}^{-1}\sigma_{ik}-{8\sqrt{2}\over 5P} \cdot
\mu^{\hbox{\scriptsize{H.S.}}}_{0}\Delta _{ik}; \nonumber \\
&&\beta_{ik}=\tilde{s}^{-1}{8\sqrt{2} \over 5P}\cdot
\mu^{\hbox{\scriptsize{H.S.}}}_{0}\Delta_{ik}-\tilde{r}^{-1}\sigma_{ik}.
\end{eqnarray}
It is clear from the analysis of  distribution  functions  of the
second  quasiequilibrium approximations   of   the   second
hydrodynamic chain that in the Grad moment method the function
$\Phi (c^{2})$ is  substituted by a constant. Finally, let us note
the simplest consequence of the variability of function $\Phi
(c^{2})$. If $\mu _{0}$ is multiplied with a small parameter
(Knudsen number $Kn$ equal to the ratio of the main free path to
the characteristic spatial scale of variations  of hydrodynamic
values),  then  the first with respect to $Kn$ approximation of
collision stress tensor $\Delta ^{(0)}_{ik}$ has the form,
\begin{equation}
\Delta ^{(0)}_{ik}=P \left( \partial _{i}u_{k}+\partial
_{k}u_{i}-{2\over 3}\delta _{ik}\partial _{s}u \right)
\label{Tri5.7}\end{equation} \noindent for Maxwell molecules, and
\begin{equation}
\Delta ^{(0)}_{ik}={5\tilde{r}\over 8\sqrt{2}
\tilde{s}\tilde{a}_{0}}P \left( \partial _{i}u_{k}+\partial
_{k}u_{i}-{2\over 3}\delta _{ik}\partial _{s}u_s \right)
\label{Tri5.8}\end{equation}

\noindent for hard spheres. Substitution  of  these  expressions
into  the momentum equations results in the Navier-Stokes equations
with effective viscosity coefficients $\mu _{\rm eff}$,
\begin{equation}
\mu _{\rm eff}=\mu ^{\hbox{\scriptsize{M.M.}}}_{0}
\label{Tri5.9}\end{equation}

\noindent for Maxwell molecules and
\begin{equation}
\mu _{\rm eff}=\tilde{a}^{-1}_{0}\mu ^{\hbox{\scriptsize{H.S.}}}_{0}
\label{Tri5.10}\end{equation}

\noindent for hard spheres. When using ten-moment Grad approximation
which does not distinguish Maxwell molecules and hard spheres,  we
obtain $\mu _{\rm eff}=\mu ^{\hbox{\scriptsize{H.S.}}}_{0}$. Some
consequences of this fact are studied below in
Sect.~\ref{moldimest}.

\section{Alternative Grad equations
and  a ``new determination of molecular dimensions" (revisited)}
\label{moldimest}

\subsection{Nonlinear functionals instead of moments  in the
closure problem}

Here we apply the method developed in the previous sections to a
classical problem: determination of molecular dimensions (as
diameters of equivalent hard spheres) from experimental viscosity
data. Scattering rates (moments of collision integral) are treated
as new independent variables, and as an alternative to moments of
the distribution function, to describe the rarefied gas near local
equilibrium. A  version of entropy maximum principle is used to
derive the   Grad-like description in terms of a finite number of
scattering rates. New equations are compared to the Grad moment
system in the heat non-conductive case. Estimations for hard
spheres demonstrate, in particular, some 10$\%$ excess of the
viscosity coefficient resulting from the scattering rate
description, as compared to the Grad moment estimation. All
necessary details of the second chain formalism that are necessary
for this example are explained below.

Here we consider a   new system of non-hydrodynamic variables,
{\it scattering rates} $M_Q(f)$:
\begin{eqnarray}
\label{MSc} M_{Q \, i_1 i_2 i_3}(f)&=&\int \mu_{i_1 i_2 i_3} Q(f,f)
\,\D \vv; \\ \nonumber \mu_{i_1 i_2
i_3}&=&mv_1^{i_1}v_2^{i_2}v_3^{i_3},
\end{eqnarray}
which, by definition, are the moments of the Boltzmann collision
integral $Q (f,f)$:
\begin{eqnarray*}
Q(f,f)=\int w(\vv^{\prime},\vv_1^{\prime},\vv,\vv_1)
\left\{f(\vv^{\prime})f(\vv_1^{\prime}) - f(\vv)f(\vv_1) \right\}
\,\D \vv^{\prime} \,\D \vv_1^{\prime} \,\D \vv_1.
\end{eqnarray*}

Here $w$ is the probability density of a change of the velocities,
$(\vv,\vv_1)\rightarrow (\vv^{\prime},\vv^{\prime}_1)$, of the two
particles after their encounter, and $w$ is defined by a model of
pair interactions. The description in terms of the scattering rates
$M_Q$ (\ref{MSc}) is alternative to the usually treated description
in terms of the moments $M$: $M_{i_1 i_2 i_3}(f)=\int \mu_{i_1 i_2
i_3} f  \,\D \vv$.

A reason to consider scattering rates instead of the moments is that
$M_Q$ (\ref{MSc}) reflect features of the interactions because of
the  $w$ incorporated in their definition, while the moments do not.
For this reason we can expect that, in general, a description with a
{\it finite} number of scattering rates will be more informative
than a description provided by the same number of their moment
counterparts.

To come to  the Grad-like equations in terms of the scattering
rates, we have to complete the following two steps: (i) to derive
a hierarchy of transport equations for $\rho$, $\uu$, $P$, and
$M_{Q \, i_1 i_2 i_3}$ in a neighborhood of the local Maxwell
states $f_0 (\rho,\uu,P)$; (ii) to truncate this hierarchy, and to
come to a closed set of equations with respect to $\rho$, $\uu$,
$P$, and a finite number of scattering rates.

In the step (i), we derive a description with infinite number of
variables, which is formally equivalent both to the Boltzmann
equation near the local equilibrium, and to the description with
an infinite number of moments. The approximation comes into play
in the step (ii) where we reduce the description to a finite
number of variables. The difference between the moment and the
alternative description occurs at this point.

The  program of two steps, (i) and (ii), is similar to what is
done in the Grad method \cite{Grad}, with the only exception (and
this is important) that we should always use scattering rates as
independent variables and not to expand them into series in
moments. Consequently, we use a method of a closure in the step 2
that does not refer to the moment expansions. Major steps of the
computation will be presented below.

\subsection{Linearization}

To complete the step (i), we represent $f$ as $f_0 (1+\varphi)$,
where $f_0$ is the local Maxwellian, and we linearize the scattering
rates (\ref{MSc}) with respect to $\varphi$:
\begin{eqnarray}
\label{DM} \Delta M_{Q\,i_1 i_2 i_3}(\varphi)&=&\int \Delta \mu_{Q
\, i_1 i_2 i_3} f_0 \varphi  \,\D \vv;
\\\nonumber \Delta \mu_{Q\,i_1 i_2 i_3}&=& L_Q ( \mu_{i_1 i_2 i_3} ).
\end{eqnarray}

Here $L_Q$ is the usual linearized collision integral, divided by
$f_0$. Though $\Delta M_Q$ are linear in $\varphi$, they are not
moments because  their microscopic densities, $\Delta \mu_Q$, are
not velocity polynomials  for a general case of $w$.

It is not difficult to derive the corresponding hierarchy of
transport equations for variables $\Delta M_{Q \, i_1 i_2 i_3}$,
$\rho$, $\uu$, and $P$  (we refer further  to this hierarchy as to
the alternative chain): one has to calculate the time derivative of
the scattering rates (\ref{MSc}) due to the Boltzmann equation, in
the linear approximation (\ref{DM}), and to complete the system with
the five known balance equations for the hydrodynamic moments
(scattering rates of the hydrodynamic moments are equal to zero due
to conservation laws). The structure of the alternative chain is
quite similar to that of the usual moment transport chain, and for
this reason we do not reproduce it here (details of calculations can
be found in \cite{book}). One should only keep in mind that the
stress tensor and the heat flux vector in the balance equations for
$\uu$ and $P$ are no more independent variables, and they are
expressed in terms of $\Delta M_{Q \, i_1 i_2 i_3}$, $\rho$, $\uu$,
and $P$.

\subsection{Truncating the chain}

To truncate the alternative chain (step (ii)), we have, first, to
choose a finite set of "essential" scattering rates (\ref{DM}), and,
second, to obtain the distribution functions which depend
parametrically only on $\rho$, $\uu$, $P$, and on the chosen set of
scattering rates. We will restrict our consideration to a single
non-hydrodynamic variable, $\sigma_{Q \, ij}$, which is the
counterpart of the stress tensor $\sigma_{ij}$. This choice
corresponds to the polynomial $mv_i v_j $ in the expressions
(\ref{MSc}) and (\ref{DM}), and the resulting equations will be
alternative to the 10 moment Grad system \footnote{To get the
alternative to the 13 moment Grad equations, one should take into
account the scattering counterpart of the  heat flux, $q_{Q \, i} =
m \int v_i \frac {v^2}{2} Q(f,f)  \,\D \vv$.}. For a spherically
symmetric interaction, the expression for $\sigma_{Q \, ij}$ may be
written:
\begin{eqnarray}
\label{es} \sigma_{Q\, ij}(\varphi)&=&\int \Delta \mu_{Q \, ij} f_0
\varphi  \,\D \vv;\\\nonumber \Delta \mu_{Q \, ij}=
 L_Q (mv_i v_j )&=&\frac {P}{\eta_{Q \, 0} (T)}S_Q (c^2)\left\{c_i c_j - \frac
 {1}{3}\delta_{ij}c^2\right\}.
\end{eqnarray}

Here $\eta_{Q \, 0} (T)$ is the first Sonine polynomial
approximation of the Chap\-man-Enskog viscosity coefficient (VC)
\cite{Chapman}, and, as usual, ${\bf c}=\sqrt {\frac {m}{2kT}}
(\vv-\uu)$. The scalar dimensionless function $S_Q$ depends only on
$c^2$, and its form depends on the  choice of interaction $w$.

\subsection{Entropy maximization}

Next, we find the functions $$f^* (\rho,\uu,P,\sigma_{Q \, ij})=f_0
(\rho,\uu,P)(1+ \varphi^* (\rho,\uu,P,\sigma_{Q \, ij}))$$ which
maximize the Boltzmann entropy $S(f)$ in a neighborhood of $f_0$
(the quadratic approximation to the entropy is valid within the
accuracy of our consideration), for fixed values of $\sigma_{Q \,
ij}$. That is, $\varphi^*$ is a solution to the following
conditional variational problem:
\begin{eqnarray}
&&\label{var} \Delta S(\varphi )  =  -\frac {k_{\rm B}}{2}\int f_0
\varphi^2  \,\D \vv\rightarrow \mbox{max},\\&& \nonumber {\rm i)}
\int \Delta \mu_{Q \, ij} f_0 \varphi  \,\D \vv=\sigma_{Q \, ij};
\quad {\rm ii)} \int \left\{1,\vv,v^2 \right\} f_0 \varphi  \,\D
\vv=0.
\end{eqnarray}
The second (homogeneous) condition in (\ref{var}) reflects that a
deviation $\varphi$ from the state $f_0$ is due only to
non-hydrodynamic degrees of freedom, and it is straightforwardly
satisfied for $\Delta \mu_{Q \, ij}$ (\ref{es}).

Notice, that if we turn to the usual moment description, then
condition (i) in (\ref{var}) would fix the stress tensor
$\sigma_{ij}$ instead of its scattering counterpart $\sigma_{Q \,
ij}$. Then the resulting function $f^* (\rho,\uu,P, \sigma_{ij})$
will be exactly the 10 moment Grad approximation. It can be shown
that a choice of any finite set of higher moments as the constraint
(i) in (\ref{var}) results in the corresponding Grad approximation.
In that sense our method of constructing $f^*$ is a direct
generalization of the Grad method onto the alternative description.

The Lagrange multipliers method gives straightforwardly the solution
to the problem (\ref{var}). After   the alternative chain is closed
with the functions $f^* (\rho,\uu,P,\sigma_{Q \, ij})$, the step
(ii) is completed, and we arrive at  a set of equations with respect
to the variables $\rho$, $\uu$, $P$, and $\sigma_{Q \, ij}$.
Switching to the variables $\zeta_{ij} = n^{-1} \sigma_{Q \, ij}$,
we have:
\begin{eqnarray}
&&\partial_t n + \partial_i (nu_i)=0;\label{a} \\ &&\rho (\partial_t
u_k + u_i \partial_i u_k )+
\partial_k P +
\partial_i \left\{ \frac {\eta_{Q \, 0} (T) n}{2r_Q P} \zeta_{ik} \right\}=0;
\label{b}\\ &&\frac {3}{2} (\partial_t P + u_i \partial_i P)+\frac
{5}{2} P \partial_i u_i + \left\{ \frac {\eta_{Q \,0} (T) n}{2r_Q P}
\zeta_{ik} \right\}\partial_i u_k =0;\label{c}\\ &&\partial_t
\zeta_{ik} + \partial_s (u_s \zeta_{ik}) +\{\zeta_{ks} \partial_s
u_i + \zeta_{is} \partial_s u_k  - \frac {2}{3} \delta_{ik}
\zeta_{rs}
\partial_s u_r \}\\ \nonumber && +  \left\{ \gamma_Q - \frac {2\beta_Q}{r_Q}\right\}\zeta_{ik}\partial_s
u_s - \frac {P^2}{\eta_{Q \, 0} (T)n}(\partial_i u_k + \partial_k
u_i -  \frac {2}{3} \delta_{ik}
\partial_s u_s ) \\ \nonumber && - \frac{\alpha_Q P}{r_Q \eta_{Q \, 0} (T)}\zeta_{ik} =0\label{d}.
\end{eqnarray}
Here $\partial_t = \partial /\partial t, \partial_i = \partial /
\partial x_i$, summation in two repeated indices is assumed, and the
coefficients $r_Q$, $\beta_Q$, and $\alpha_Q$ are defined with the
help of the  function $S_Q$ (\ref{es}) as follows:
\begin{eqnarray}
\label{constants} r_Q &=& \frac {8}{15\sqrt{\pi}}\int_0^{\infty}
\mbox{e}^{-c^2}c^6 \left( S_Q (c^2) \right)^2  \,\D c; \nonumber \\
\nonumber \beta_Q &=& \frac {8}{15\sqrt{\pi}}\int_0^{\infty}
\mbox{e}^{-c^2}c^6 S_Q (c^2)\frac{ \,\D S_Q (c^2)}{ \,\D (c^2)}
\,\D c; \\  \alpha_Q &=& \frac {8}{15\sqrt{\pi} }\int_0^{\infty}
\mbox{e}^{-c^2}c^6 S_Q (c^2)R_Q (c^2)  \,\D c.
\end{eqnarray}
The function $R_Q (c^2)$ in the last expression is defined due to
the action of the operator $L_Q$ on the function $S_Q (c^2)(c_i c_j
- \frac{1}{3}\delta_{ij}c^2 )$:
\begin{equation}
\label{action} \frac{P}{\eta_{Q \, 0}}R_Q (c^2)(c_i c_j - \frac
{1}{3}\delta_{ij}c^2)= L_Q (S_Q (c^2)(c_i c_j -
\frac{1}{3}\delta_{ij}c^2 )).
\end{equation}
Finally, the parameter $\gamma_Q$ in (\ref{a}-\ref{d}) reflects the
temperature dependence of the VC:
\[
\gamma_Q = \frac {2}{3}\left( 1- \frac {T}{\eta_{Q \,0} (T)} \left(
\frac {\D \eta_{Q \,0} (T)}{\D T} \right) \right).
\]
The set of ten equations (\ref{a}-\ref{d}) is alternative to the 10
moment Grad equations.

\subsection{A new determination of molecular dimensions (revisited)}

The observation already made is that for Maxwell molecules we have:
$S^{{\rm M.M.}} \equiv 1$, and $\eta_0^{{\rm M.M.}} \propto T$; thus
$\gamma^{{\rm M.M.}}= \beta^{{\rm M.M.}}=0$, $r^{{\rm
M.M.}}=\alpha^{{\rm M.M.}}=\frac{1}{2}$, and (\ref{a}-\ref{d})
becomes the 10 moment Grad system under a simple change of variables
$ \lambda\zeta_{ij}=\sigma_{ij}$, where $\lambda$ is the
proportionality coefficient in the temperature dependence of
$\eta_0^{{\rm M.M.}}$.

These properties (the function $S_Q$ is a constant, and the VC is
proportional to $T$) are true only for Maxwell molecules. For all
other interactions, the function $S_Q$ is not identical to one, and
the VC $\eta_{Q \, 0} (T)$ is not proportional to $T$. Thus, the
shortened alternative description is not equivalent indeed to the
Grad moment description. In particular, for hard spheres, the exact
expression for the function $S^{{\rm H.S.}}$ (\ref{es}) reads:
\begin{eqnarray}
\label{RS} S^{{\rm H.S.}} &=& \frac {5\sqrt 2}{16} \int_0^1 \exp
(-c^2 t^2)(1-t^4 ) \left( c^2 (1-t^2 ) + 2 \right)  \,\D t;
\\\nonumber \eta_0^{{\rm H.S.}} &\propto& \sqrt T.
\end{eqnarray}

Thus, $\gamma^{{\rm H.S.}}=\frac {1}{3}$, and $\frac{\beta^{{\rm
H.S.}}}{r^{{\rm H.S.}}}\approx 0.07$, and the equation for the
function $\zeta_{ik}$  (\ref{d}) contains a nonlinear term,
\begin{equation}
\label{nonlin} \theta^{{\rm H.S.}}\zeta_{ik}\partial_s u_s ,
\end{equation}
where $\theta^{{\rm H.S.}}\approx 0.19$. This term is missing in the
Grad 10 moment equation.

Finally, let us  evaluate the  VC which results from the alternative
description (\ref{a}-\ref{d}). Following Grad's arguments
\cite{Grad}, we see that, if the relaxation of $\zeta_{ik}$ is fast
compared to the hydrodynamic variables, then the two last terms in
the equation for $\zeta_{ik}$ (\ref{a}-\ref{d}) become dominant, and
the equation for $\uu$ casts into the standard Navier-Stokes form
with an effective VC $\eta_{Q \, {\rm eff}}$:
\begin{equation}
\label{effective} \eta_{Q \, {\rm eff}} = \frac {1}{2\alpha_Q}
\eta_{Q \, 0} .
\end{equation}

For Maxwell molecules, we easily derive that the coefficient
$\alpha_Q$ in (\ref{effective}) is equal to $\frac{1}{2}$. Thus,
as one expects, the effective VC (\ref{effective}) is equal to the
Grad value, which, in turn, is equal to the exact value in the
frames of the Chapman--Enskog method for this model.

For all interactions different from the Maxwell molecules, the VC
$\eta_{Q \, {\rm eff}}$ (\ref{effective}) is not equal to $\eta_{Q
\, 0} $. For hard spheres, in particular, a computation of the VC
(\ref{effective}) requires  information about the function $R^{{\rm
H.S.}}$ (\ref{action}). This is achieved upon a substitution of the
function $S^{{\rm H.S.}}$ (\ref{RS}) into (\ref{action}). Further,
we have to  compute  the action of the operator $L^{{\rm H.S.}}$ on
the function $S^{{\rm H.S.}}(c_i c_j - \frac{1}{3}\delta_{ij}c^2 )$,
which is rather complicated. However, the VC $\eta^{{\rm
H.S.}}_{{\rm eff}}$ can be relatively easily estimated  by using a
function $S^{{\rm H.S.}}_a = \frac {1}{\sqrt 2} (1 + \frac {1}{7}
c^2 )$, instead of the function $S^{{\rm H.S.}}$, in (\ref{action}).
Indeed, the function $S^{{\rm H.S.}}_a$ is tangent to the function
$S^{{\rm H.S.}}$ at $c^2 = 0$, and is its majorant (see Fig.\
\ref{Hard}). Substituting $S^{{\rm H.S.}}_a$ into (\ref{action}),
and computing the action of the collision integral, we find the
approximation $R_a^{{\rm H.S.}}$; thereafter we evaluate the
integral $\alpha^{{\rm H.S.}} $ (\ref{constants}), and finally come
to the following expression:

\begin{figure}[t]
\begin{centering}
\includegraphics[width=73mm, height=53mm]{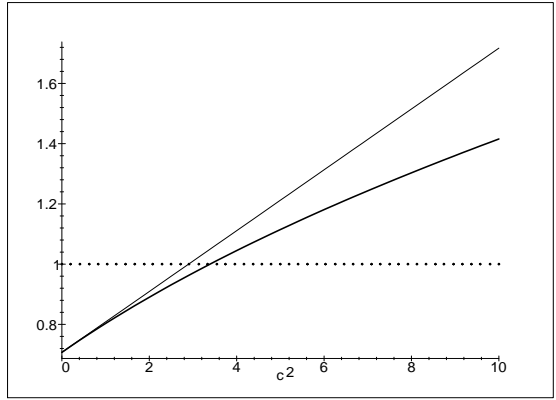}
\caption {\label{Hard} Approximations for hard spheres:
    bold line -- function $S^{{\rm H.S.}}$,
    solid line -- approximation $S_a^{{\rm H.S.}}$,
    dotted line -- Grad moment approximation.}
    \end{centering}
\end{figure}

\begin{equation}
\label{visc} \eta^{{\rm H.S.}}_{{\rm eff}} \ge
\frac{75264}{67237}\eta_0^{{\rm H.S.}} \approx 1.12\eta_0^{{\rm
H.S.}}.
\end{equation}

Thus, for hard spheres, the description in terms of scattering rates
results in  the VC of  more than 10$\%$ higher than in the Grad
moment description.

A discussion of the results concerns the following two items.

1. Having two not equivalent descriptions which were obtained within
one method, we may ask: which is more relevant? A simple test is to
compare characteristic times of an approach to hydrodynamic regime.
We have $\tau_G \sim \eta_0^{{\rm H.S.}}/P$ for 10-moment
description, and $\tau_a \sim \eta^{{\rm H.S.}}_{{\rm eff}}/P$ for
alternative description. As $\tau_a > \tau_G $, we see that
scattering rate decay slower than corresponding moment, hence, at
least for rigid spheres, the alternative description is more
relevant. For Maxwell molecules both the descriptions are, of
course, equivalent.

\begin{table}
\caption{\label{TabVir}Three virial coefficients: experimental
$B_{{\rm exp}}$, classical $B_0 $ \cite{Hirschfelder}, and reduced
$B_{{\rm eff}}$ for three gases at $T=500 K$}
\begin{center}
\begin{tabular}{|c|c|c|c|}\hline
&$B_{{\rm exp}}$&$B_0 $&$B_{{\rm eff}}$ \\ \hline Argon&8.4&60.9&50.5\\ Helium&10.8&21.9&18.2\\
Nitrogen &168&66.5&55.2\\ \hline
\end{tabular}
\end{center}
\end{table}

2. The VC $\eta^{{\rm H.S.}}_{{\rm eff}} $ (\ref{visc}) has the same
temperature dependence as $\eta_0^{{\rm H.S.}} $, and also the same
dependence on a scaling parameter (a diameter of the sphere). In the
classical book \cite{Chapman} (pp.\ 228-229), "sizes" of molecules
are presented, assuming that a molecule is represented with an
equivalent sphere and VC is estimated as $\eta_0^{{\rm H.S.}} $.
Since our estimation of VC differs only by a dimensionless factor
from $\eta_0^{{\rm H.S.}} $, it is straightforward to conclude that
effective sizes of molecules will be reduced by the factor $b$,
where $$b= \sqrt{\eta_0^{{\rm H.S.}}/ \eta^{{\rm H.S.}}_{{\rm eff}}
} \approx 0.94 .$$ Further, it is well known that sizes of molecules
estimated via viscosity in \cite{Chapman} disagree with the
estimation via the virial expansion of the equation of state. In
particular, in book \cite{Hirschfelder}, p.\ 5, the measured second
virial coefficient $B_{{\rm exp}}$ was compared with the calculated
$B_0 $, in which the diameter of the sphere was taken from the
viscosity data. The reduction of the diameter  by factor $b$ gives
$B_{{\rm eff}}=b^3 B_0 $. The values $B_{{\rm exp}}$ and $B_0 $
\cite{Hirschfelder} are compared with $B_{{\rm eff}}$ in the Table
\ref{TabVir} for three gases at $T=500 K$. The results for argon and
helium are better for $B_{{\rm eff}}$, while for nitrogen $B_{{\rm
eff}}$ is worth than $B_0 $. However, both  $B_0$ and $B_{{\rm
eff}}$ are far from the experimental values.

Hard spheres is, of course, an oversimplified model of
interaction, and the comparison presented does not allow for a
decision between $\eta_0^{{\rm H.S.}}$ and  $ \eta^{{\rm
H.S.}}_{{\rm eff}}$. However, this simple example illustrates to
what extend the correction to the VC can affect a comparison with
experiment. Indeed, as it is well known, the first-order Sonine
polynomial computation for the Lennard-Jones (LJ) potential gives
a very good fit of the temperature dependence of the VC for all
noble gases \cite{Dorf}, subject to a proper choice of the two
unknown scaling parameters of the LJ potential \footnote{A
comparison of molecular parameters of the LJ potential, as derived
from the viscosity data, to those obtained from independent
sources, can be found elsewhere, e.g. in \cite{Chapman}, p.\
237.}. We may expect that a dimensionless correction of the VC for
the LJ potential might be of the same order as above for rigid
spheres. However, the functional character of the temperature
dependence will not be affected, and a fit will be obtained
subject to a different choice of the molecular parameters of the
LJ potential. The five--parametric family of pair potentials was
discussed in Ref. \cite{Kestin}. These five constants for each
pair potential have been determined by a fit to experimental data
with some additional input from theory. After that, the
Chapman--Enskog formulas for the second virial coefficient and
main transport coefficients give satisfactory description of
experimental data    \cite{Kestin}. Such a semiempirical approach
that combines fitting with kinetic theory might be very successful
in experimental data description, but does not allow us to make a
choice between hierarchies. This choice requires less flexibility
in the potential construction. The best solution here is
independent determination of the interaction potential without
references to transport coefficients or thermodynamic data.

\section*{Appendix: Formulas  of the second
quasiequilibrium approximation  of the second and mixed
hydrodynamic chains  for Maxwell molecules and hard spheres}

Let us write $\nu _{Qik}$ (\ref{Tri4.2}) in the standard form:
\begin{eqnarray}
&&\nu _{Qik}\\ &&=\int f^{(0)} \mid {\vv_1}-{\vv} \mid \left\{
(v_{i}-u_{i})(v_{k}-u_{k})-{1\over 3}\delta
_{ik}({\vv}-{\uu})^{2}\right\} b  \,\D b  \,\D \epsilon  \,\D
{\vv_1}, \nonumber \label{TriA.1}
\end{eqnarray}
\noindent where $b$ is the impact parameter, $\epsilon $ is the
angle between  the  plane containing the trajectory of the
particle being scattered  in  the system of the center of mass and
the plane containing the entering asymptote, the trajectory,  and
a certain fixed   direction.   It   is convenient to switch to the
dimensionless velocity ${\bf c}$:
\begin{equation}
c_{i}=\left({m\over 2k_{\rm B}T} \right)^{1/2} (v_{i}-u_{i})
\label{TriA.2}
\end{equation}

\noindent and to the dimensionless relative velocity ${\bf g}$:
\begin{equation}
g_{i}={1\over 2} \left({m\over k_{\rm B}T} \right)^{1/2}
(v_{1i}-u_{i}) \label{TriA.3}
\end{equation}

\noindent After standard transformations and integration with
respect  to $\epsilon$ (see \cite{Chapman}) we obtain in
(\ref{TriA.1})
\begin{eqnarray}
\nu _{Qik}&=&{3P\over m}\pi ^{-1/2} \\  &\times&   \int \exp
(-c^{2}_{1})\varphi ^{(2)}_{1}(g) \left(
(c_{1i}-c_{i})(c_{1k}-c_{k})-{1\over 3}\delta _{ik}({\bf
c}_{1}-{\bf c})^{2} \right)  \,\D {\bf c_1}. \nonumber
\label{TriA.4}
\end{eqnarray}

\noindent Here
\begin{equation}
\varphi^{(2)}_{1}=\int (1-\cos^{2}\chi )\mid {\vv_1}-{\vv}\mid
b(\chi ) \biggl|{ \,\D b \over
 \,\D \chi} \biggr| d\chi, \label{TriA.5}
\end{equation}

\noindent and $\chi $ is an angle between the vectors $g$ and
$g'$.

The dependence of $\varphi ^{(2)}_{1}$ on the vector $g$ is
determined  by  the choice of  the  model of  particle's
interaction.

For  Maxwell molecules,
\begin{equation}
\varphi^{(2)}_{1}=\left( {2\kappa \over m} \right)^{1/2} A_{2}(5),
\label{TriA.6}
\end{equation}

\noindent where $\kappa $ is a force constant, $A_{2}(5)$ is a
number, $A_{2}(5)\approx 0.436.$

For the model of hard spheres
\begin{equation}
\varphi ^{(2)}_{1}={\sqrt{2} \sigma^{2}\over 3} \left({k_{\rm
B}T\over m}\right)^{1/2}\mid {\bf c_1}-{\bf c}\mid, \label{TriA.7}
\end{equation}

\noindent where $\sigma $ is diameter of the sphere modelling the
particle.

Substituting (\ref{TriA.6}) and (\ref{TriA.7}) into
(\ref{TriA.4}), we transform  the  latter to the form:

\noindent for Maxwell molecules
\begin{eqnarray}
&&\nu _{Qik}={3P\over 4m} \left( {2\kappa \over \pi m}
\right)^{1/2} A_{2}(5)\exp (-c^{2}) \left({\partial\over \partial
c_{i}}{\partial \over \partial c_{k}}-{1\over 3}\delta
_{ik}{\partial \over \partial c_{s}}{\partial \over \partial c_s}
\right) T^{\hbox{\scriptsize{M.M.}}}(c^{2}); \nonumber
\\ && T^{\hbox{\scriptsize{M.M.}}}(c^{2})=\int \exp (-x^{2}-2x_{k}c_{k}) \,\D {\xx}; \label{TriA.8}
\end{eqnarray}

\noindent for hard spheres
\begin{eqnarray}
&&\nu _{Qik}={P\sigma ^{2}\over 2\sqrt{2} m}\left({k_{\rm B}T\over
\pi m}\right)^{1/2} \exp (-c^{2}) \left({\partial\over \partial
c_{i}}{\partial \over \partial c_{k}}- {1\over 3}\delta
_{ik}{\partial \over \partial c_{s}}{\partial \over \partial
c_s}\right) T^{\hbox{\scriptsize{H.S.}}}(c^{2}); \nonumber
\\ && T^{\hbox{\scriptsize{H.S.}}}(c^{2})=
\int \mid {\xx}\mid \exp (-x^{2}-2x_{k}c_{k}) \,\D {\xx}.
\label{TriA.9}
\end{eqnarray}

It is an easy matter to perform  integration  in  (\ref{TriA.8}),
the integral is equal to $\pi ^{3/2}e^{c^{2}}$.

\noindent Therefore for Maxwell molecules,
\begin{equation}
\nu _{Qik}={3\over 2}n\pi \left({2\kappa \over m}\right)^{1/2}
A_{2}(5) \left( (v_{i}-u_{i})(v_{k}-u_{k})-{1\over 3}\delta
_{ik}({\vv}-{\uu})^{2} \right). \label{TriA.10}
\end{equation}

\noindent The integral $T^{\hbox{\scriptsize{H.S.}}}$ in
(\ref{TriA.9}) can be transformed as follows:
\begin{equation}
T^{\hbox{\scriptsize{H.S.}}}(c^{2})=2\pi +\pi \int^{+1}_{-1} \exp(
c^{2}(1-y^{2})) c^{2}(1+y^{2}) \,\D y. \label{TriA.11}
\end{equation}

\noindent Then for the model of hard spheres,
\begin{eqnarray}
\nu _{Qik}&=&\sqrt{2 \pi} n\sigma ^{2} \left({k_{\rm B}T\over
m}\right)^{3/2} \left( c_{i}c_{k}-{1\over 3}\delta _{ik}c^{2}
\right) \nonumber \\ & \times &  \int^{+1}_{-1} \exp (-c^{2}y^{2})
(1+y^{2})(1-y^{2})(c^{2}(1-y^{2})+2)  \,\D y. \label{TriA.12}
\end{eqnarray}

\noindent Let us note a useful relationship:
\begin{eqnarray}
&& d^{n}T^{\hbox{\scriptsize{H.S.}}}/d(c^{2})^{n}=\pi
\int^{+1}_{-1} \exp (c^{2}(1-y^{2})) \nonumber
\\ && \; \; \times  (1+y^{2})(1-y^{2})^{n-1}(c^{2}(1-y^{2})+n) \,\D y, n\geq 1. \label{TriA.13}
\end{eqnarray}

Use the expressions for the viscosity  coefficient $\mu _{0}$
which are obtained in the  first approximation  of  the
Chapman--Enskog method:

\noindent for Maxwell molecules,
\begin{equation}
\mu ^{\hbox{\scriptsize{M.M.}}}_{0}=\left( {2m\over \kappa }
\right) ^{1/2} {k_{\rm B}T\over 3\pi A_{2}(5)}; \label{TriA.14}
\end{equation}

\noindent for hard spheres,
\begin{equation}
\mu^{\hbox{\scriptsize{H.S.}}}_{0}={5(k_{\rm B}Tm)^{1/2}\over
16\pi ^{1/2}\sigma ^{2}}. \label{TriA.15}
\end{equation}

Transformation of (\ref{TriA.10}), (\ref{TriA.12}) to the form of
(\ref{Tri4.5}) gives the following functions $\Phi
(({\vv}-{\uu})^{2})$:

\noindent for Maxwell molecules,
\begin{equation}
\Phi =P / \mu^{\hbox{\scriptsize{M.M.}}}_{0}; \label{TriA.16}
\end{equation}

\noindent for hard spheres
\begin{eqnarray}
\Phi& =&{5P\over 16 \sqrt{2} \mu ^{\hbox{\scriptsize{H.S.}}}_{0}}
\int^{+1}_{-1} \exp \left(-{m({\vv}-{\uu})^{2}\over 2k_{\rm
B}T}y^{2} \right) \nonumber \\ & \times & (1+y^{2})(1-y^{2})
\left({m({\vv}-{\uu})^{2}\over 2k_{\rm B}T}(1-y^{2})+2 \right) dy.
\label{TriA.17}
\end{eqnarray}

The parameter $r$ from (\ref{Tri4.6}) is:

\noindent for Maxwell molecules:
\begin{eqnarray}
r=\left( m\mu ^{\hbox{\scriptsize{M.M.}}}_0\right)^2
/(2P^{3}k_{\rm B}T); \label{TriA.18}
\end{eqnarray}

\noindent for hard spheres:
\begin{equation}
r=\tilde{r}{64 \left(m \mu^{\hbox{\scriptsize{M.M.}}}_0\right)^2
\over 25P^{3}k_{\rm B}T}. \label{TriA.19}
\end{equation}

\noindent The dimensionless  parameter $\tilde{r}$  is represented
as follows:
\begin{eqnarray}
\tilde{r}^{-1}&=&{1\over 16} \int^{+1}_{-1} \int^{+1}_{-1}
\alpha^{-11/2}\beta (y)\beta (z)\gamma (y)\gamma (z)\nonumber \\ &
\times & (16\alpha ^{2}+28\alpha (\gamma (y)+\gamma (z))+63\gamma
(y)\gamma (z))  \,\D y \,\D z. \label{TriA.20}
\end{eqnarray}

\noindent Here and below the following notations are used:
\begin{equation}
\beta (y)=1+y^{2},\qquad  \gamma (y)=1-y^{2},\qquad \alpha
=1+y^{2}+z^{2}. \label{TriA.21}
\end{equation}

\noindent Numerical value of $\tilde{r}^{-1}$ is 5.212 to third
decimal point.

The parameter (\ref{Tri4.14}) is:

\noindent for Maxwell molecules
\begin{equation}
s^{-1}=(2P^{2}k_{\rm B}T)/ \left( m^{2}\mu
^{\hbox{\scriptsize{M.M.}}}_0 \right); \label{TriA.22}
\end{equation}

\noindent for hard spheres
\begin{equation}
s^{-1}=\tilde{s}^{-1}{5\sqrt{2} P^{2}k_{\rm B}T\over 8m^{2}\mu
^{\hbox{\scriptsize{H.S.}}}_{0}}. \label{TriA.23}
\end{equation}

\noindent The dimensionless parameter $\tilde{s}^{-1}$ is of the
form
\begin{equation}
\tilde{s}^{-1}= \int^{+1}_{-1} \gamma (y)\beta ^{-7/2}(y) \left(
\beta (y)+{7\over 4}\gamma (y) \right)  \,\D y. \label{TriA.24}
\end{equation}

\noindent Numerical value of $\tilde{s}^{-1}$ is 1.115 to third
decimal point.

The scattering rate density (\ref{Tri4.21}) is of the form,
\begin{equation}
\xi_{Qi}=\sqrt{2} \left( {k_{\rm B}T\over m} \right)^{3/2} \int
f^{(0)} ( {\vv}_{1})\mid {\vv_1}-{\vv}\mid \left\{ c_i \left(
c^{2}-{5\over 2} \right) \right\}  b \,\D b \,\D  \epsilon
d{\vv_1}. \label{TriA.25}
\end{equation}

\noindent Standard  transformation  of  the  expression $\left\{
c_{i}(c^{2}-5/2)\right\} $   and integration with respect to
$\epsilon $ change (\ref{TriA.25}) to the form,
\begin{eqnarray}
&&\xi _{Qi}\\&& ={P\over \sqrt{2 \pi} m} \int \exp
(-c^{2}_{1})\varphi
^{(2)}_{1}(3(c^{2}_{1}-c^{2})(c_{1i}-c_{i})-({\bf c}_{1}-{\bf
c})^{2}(c_{1i}+c_{i})) \,\D {\bf c_1}. \nonumber \label{TriA.26}
\end{eqnarray}

\noindent Further, using the  expressions  (\ref{TriA.6})  and
(\ref{TriA.7})  for $\varphi ^{(2)}_{1}$, we obtain:

\noindent for Maxwell molecules:
\begin{equation}
\xi _{Qi}={P\over m^{2}}\left( {\kappa k_{\rm B}T\over \pi
}\right) ^{1/2}A_{2}(5)\exp \left(-c^{2}\right)
\hat{D}_{i}T^{\hbox{\scriptsize{M.M.}}}(c^{2}); \label{TriA.27}
\end{equation}

\noindent for hard spheres:
\begin{equation}
\xi _{Qi}={Pk_{\rm B}T\sigma ^{2}\over \sqrt{\pi} m^{2}}\exp
(-c^{2})\hat{D}_{i}T^{\hbox{\scriptsize{H.S.}}}(c^{2}).
\label{TriA.28}
\end{equation}

\noindent The operator $\hat{D}_{i}$ is of the form
\begin{equation}
{1\over 4} {\partial \over \partial c_{i}}{\partial \over \partial
c_{s}}{\partial \over \partial c_{s}} + {3\over 2}c_{s} {\partial
\over \partial c_{s}}{\partial \over \partial c_{i}} - {1\over
2}c_{i} {\partial \over \partial c_{s}}{\partial \over \partial
c_{s}}. \label{TriA.29}
\end{equation}

\noindent The operator $\hat{D}_{i}$ acts on the function $\psi
(c^{2})$ as follows:
\begin{equation}
{d^{2}\psi \over d(c^{2})^{2}}2c_i \left( c^{2}-{5\over 2} \right)
+c_{i}c^{2} \left( {d^{2}\psi \over d(c^{2})^{2}} - {d^{3}\psi
\over d(c^{2})^{3}} \right). \label{TriA.30}
\end{equation}

\noindent From (\ref{TriA.27}), (\ref{TriA.28}) we obtain:

\noindent for Maxwell molecules:
\begin{equation}
\xi _{Qi}={P\over 3\mu
^{\hbox{\scriptsize{M.M.}}}_{0}}(v_{i}-u_{i})\left(({\vv}-{\uu})^{2}-{5k_{\rm
B}T\over m}\right); \label{TriA.31}
\end{equation}

\noindent for hard spheres:
\begin{eqnarray}
&&\xi _{Qi}={5P\over 16\sqrt{2} \mu
^{\hbox{\scriptsize{H.S.}}}_{0}}\left\{ (v_{i}-u_{i})\left(
({\vv}-{\uu})^{2}-{5k_{\rm B}T\over m}\right)  \right.  \\ &&
\times \int^{+1}_{-1} \exp \left( -{m({\vv}-{\uu})^2 \over 2k_{\rm
B}T} y^{2} \right) \beta (y)\gamma (y) \left( {m({\vv}-{\uu})\over
2k_{\rm B}T}^{2}\gamma (y)+2 \right) dy\nonumber
\\ && + (v_{i}-u_{i})({\vv}-{\uu})^{2} \nonumber  \\&& \left.
\times \int^{+1}_{-1} \exp \left(-{m({\vv}-{\uu})\over 2k_{\rm
B}T}^{2}y^{2} \right) \beta (y)\gamma (y) \left( \sigma
(y){m({\vv}-{\uu})\over 2k_{\rm B}T}^{2}+\delta (y) \right) \,\D y
\right\}. \nonumber \label{TriA.32}
\end{eqnarray}

\noindent The functions $\sigma (y), \delta (y)$ are of the form
\begin{equation}
\sigma (y)=y^{2}(1-y^{2}),\qquad  \delta (y)=3y^{2}-1.
\label{TriA.33}
\end{equation}

The parameter $\eta $ from (\ref{Tri4.25}) is:

\noindent for Maxwell molecules:
\begin{equation}
\eta ={9m^3 \left( \mu ^{\hbox{\scriptsize{M.M.}}}_0  \right)^2
\over 10P^{3}(k_{\rm B}T)^{2}}; \label{TriA.34}
\end{equation}

\noindent for hard spheres:
\begin{equation}
\eta =\tilde{\eta}  { 64m^{3} \left(\mu
^{\hbox{\scriptsize{H.S.}}}_0 \right)^2 \over 125P^{3}(k_{\rm
B}T)^{2}}. \label{TriA.35}
\end{equation}

\noindent The dimensionless parameter $\tilde{\eta }$ is of the
form
\begin{eqnarray}
\tilde{\eta }^{-1}&=& \int^{+1}_{-1} \int^{+1}_{-1} \beta (y)\beta
(z)\gamma (y) \gamma (z)\alpha^{-13/2}\left\{ {639\over 32}(\gamma
(y)\gamma (z)+\sigma (y)\sigma (z)\right. \nonumber
\\& +& \sigma (y)\gamma (z)+\sigma (z)\gamma (y))+{63\over 16}\alpha (2\gamma (y)+2\gamma (z)-5\gamma
(y)\gamma (z) \nonumber \\ & +& 2(\sigma (y)+\sigma (z))+\gamma
(z)\delta (y)+\gamma (y)\delta (z)+\sigma (y)\delta (z)+\sigma
(z)\delta (y))\nonumber \\ & + &{7\over 8}\alpha ^{2}(4-10\gamma
(y)-10\gamma (z))+{25\over 4}\gamma (y)\gamma (z)+2\delta (y) \\ &
+ &2\delta (z)-5\sigma (y)-5\sigma (z)-{5\over 2}(\gamma (z)\delta
(y)+\gamma (y)\delta (z)+\delta (y)\delta (z))\nonumber
\\ & +& \left. {1\over 4}\alpha ^{3} \left( -20+{25\over 4}(\gamma (y)+\gamma (z))-5(\delta (y)+\delta
(z)) \right)+{5\over 2}\alpha ^{4}\right\}   \,\D y \,\D z.
\nonumber \label{TriA.36}
\end{eqnarray}

\noindent Numerical value of $\tilde{\eta }^{-1}$ is 0.622 to
second decimal point.
\par
Finally, from (\ref{TriA.31}), (\ref{TriA.32}) we obtain
$\tau^{-1} (\ref{Tri4.32})$:
\par
\noindent for Maxwell molecules
\begin{equation}
\tau^{-1}={5(Pk_{\rm B}T)^{2}\over 3\mu
^{\mbox{\scriptsize{M.M.}}}_{0}m^{3}};\label{TriA.37}
\end{equation}
for hard spheres
\begin{eqnarray}\label{TriA.38}
\nonumber \tau^{-1}&=&\tilde{\tau}^{-1}{25P^{2}(k_{\rm
B}T)^{2}\over 8\sqrt{2} m^{3} \mu
^{\mbox{\scriptsize{H.S.}}}_{0}}; \\ \nonumber
\tilde{\tau}^{-1}&=&{1\over 8} \int^{+1}_{-1} \beta
^{-9/2}(y)\gamma (y) \{ 63(\gamma (y)+\sigma (y))\\ \nonumber &
+&7\beta (y)(4-10\gamma (y)+2\delta (y)-5\sigma (y))+20\beta
^{3}(y)\\ &+& \beta ^{2}(y)(25\gamma (y)-10\delta (y)-40) \}  \,\D
y \approx 4.322.
\end{eqnarray}

\section{Conclusion and outlook}

We developed the Triangle Entropy Method (TEM) for model reduction
and demonstrated how it works for the Boltzmann equation. Moments
of  the  Boltzmann  collision integral, or scattering rates are
treated  as  independent variables rather than as infinite moment
series. Three classes of reduced models are constructed. The
models of the first class involve only moments of distribution
functions, and coincide with those of the Grad method in the
Maximum Entropy version. The models of the second type involves
only scattering rates. Finally, the mixed description models
involve both the moments and the scattering rates. TEM allows us
to obtain all the closure formulas in explicit form, not only for
the Maxwell molecules (as it is usual), but for hard spheres also.
We found the new Boltzmann--kinetics estimations for the
equivalent hard spheres radius for gases.

The main benefits from TEM are:
\begin{enumerate}
\item{It constructs the closure as a solution of linear equations,
and, therefore, often gives it in an explicit form;}
\item{It provides the thermodynamic properties of reduced models, at least, locally;}
\item{It admits nonlinear functionals as macroscopic variables,
this possibility is important for creation of non-equilibrium
thermodynamics of non-linear fluxes, reaction rates, scattering
rates, etc.}
\end{enumerate}

The following fields for future TEM applications are important:
\begin{itemize}\item{Modelling of nonequilibrium processes
in gases (Boltzmann kinetics and its generalisations);}
\item{Chemical kinetics models with reaction rates as independent
variables;}
\item{Kinetics of complex media (non-Newtonian liquids, polymers, etc.) with
the Fokker--Planck equation as the basic kinetic description.}
\end{itemize}
Of course, TEM is a new computational version of the entropy
maximum method (MaxEnt) that is in wide use in the Extended
irreversible thermodynamics (EIT) \cite{EIT} after  works of
Jaynes \cite{Janes1}, Kogan and Rosonoer \cite{KoRoz,Ko,Roz} and
Lewis \cite{Lew}.

MaxEnt methods obtain the second wind because the high interest to
non-classical nonextensive entropies \cite{Tsa,Abe}. In this
sense, the Fokker--Plank equation seems to be a very attractive
example for MaxEnt method application, and, in particular, for the
application of TEM. This classical equation admits a broad class
of Lyapunov functions, including nonextensive entropies.

The Fokker--Planck equation (FPE) is a familiar model in various
problems of nonequilibrium statistical physics We consider the FPE
of the form
\begin{equation}
\label{SFP} {\partial W(\xx,t) \over \partial t} ={\partial \over
\partial \xx} \left\{D \left[W{\partial \over
\partial \xx} U +{\partial \over \partial \xx} W\right]\right\}.
\end{equation}
Here, $W(\xx,t)$ is the probability density over the configuration
space $x$ at time $t$, while $U(\xx)$ and $D(\xx)$ are the
potential and the positively semi-definite ($ (\yy,D\yy)\ge 0$)
diffusion matrix.

It is known that for the Boltzmann equation there exists only one
universal Lyapunov functional: the entropy (we do not distinguish
functionals which are related to each other by monotonic
transformations). For the FPE there exists a whole family of
universal Lyapunov functionals. Let $h(a)$ be a convex function of
one variable $a\geq 0$, $h''(a)>0$,

\begin{equation} \label{FPES}
S_h[W]=-\int W_{\rm eq}(\xx) h\left[\frac{W(\xx,t)}{W_{\rm
eq}(\xx)}\right] \,\D x.
\end{equation}

The density of production of the generalized entropy $S_h$,
$\sigma_h$, is nonnegative:

\begin{equation} \label{sigmah}
\sigma_h(\xx)=W_{\rm eq}(\xx)h''\left[\frac{W(\xx,t)}{W_{\rm
eq}(\xx)}\right]\left({\partial \over \partial
\xx}\frac{W(\xx,t)}{W_{\rm eq}(\xx)},D{\partial \over
\partial \xx}\frac{W(\xx,t)}{W_{\rm eq}(\xx)}\right)\geq 0.
\end{equation}

The most important variants for the choice of $h$ are:
\begin{itemize}
\item{$h(a)=a\ln a$, and $S_h$ is the
Boltzmann--Gibbs--Shannon entropy (in the Kullback form
\cite{Kull,Pla}),} \item{$h(a)=a\ln a-\epsilon \ln a$,
$\epsilon>0$, and $S_h^\epsilon$ is the maximal family of {\it
additive} entropies \cite{ENTR1,ENTR2,ENTR3} (these entropies are
additive for the composition of independent subsystems).}
\item{$h(a)=\frac{1-a^q}{1-q}$, and $S_h^q$ is the family of
Tsallis entropies \cite{Tsa,Abe}. These entropies are not
additive, but become additive after a nonlinear monotonous
transformation. This property can serve as a definition of the
Tsallis entropies in the class of generalized entropies
(\ref{FPES}) \cite{ENTR3}.}
\end{itemize}

The MaxEnt closure approximations for the Fokker--Planck equation
are developed for the classical Boltzmann--Gibbs--Shannon entropy
in the Kullback form and standard linear functionals (moments) as
macroscopic variables \cite{Pla,IKOePhA02,IKOePhA03}. Now we are
in position to develop the whole family of approximation in
explicit form (due to TEM), for classical and non-classical
entropies, for linear and non-linear macroscopic variables.

The crucial question is: where to stop? Is it possible to decide,
is this particular model from the hierarchy sufficiently accurate,
or we need to go ahead? Without criteria for such a decision we
have infinite number of theories.

The residual estimates are possible: we can estimate the defect of
invariance (see Fig.~\ref{fig1QEHier}). If it is too big (in
comparison with the full right hand side $J$), then we should
switch to the next system of hierarchy. If it vanishes, we could
try the previous system. Normally, it is impossible to find one
reduced model for all regimes, but it is possible to change the
model during simulation.

There exists one more benefit from the hierarchy. For each model
we have a correspondent approximate slow invariant manifold
$\Omega_i$, and a vector field of reduced dynamics $J_i$ which is
defined at points from $\Omega_i$ and is tangent to this manifold.
This structure gives a possibility to estimate not the whole
defect of invariance, but a ``partial defect" $\Delta_i =
J_{i+1}-J_i$. Usually it is sufficient to estimate this partial
defect of invariance, that is, to check whether the current model
is the approximate slow invariant manifold for the next model up
to desired accuracy. Examples of these estimates and applications
are presented in Refs.~\cite{CMIM,GKSpri,IKar2,GKIOeNONNEWT2001}.
We propose to use the flexible technology of modeling with
adaptive choice of the model from hierarchy. This approach could
be discussed as intermediate one between the classical one--model
calculations and the equation--free approach \cite{KevFree}, for
example.

We construct the quasiequilibrium hierarchy of models for a system
with entropy growth. These systems relax to equilibrium points.
But most interesting application is modeling of open systems. It
is possible to use obtained hierarchy of models for open systems
just by adding flows under the assumption that the fast motion and
slow manifold do not change due to the system opening. For
example, we usually use the Navier--Stokes equation for systems
with external flows that do not relax to equilibrium. If the
external flows are fast and the perturbation of slow manifold is
significant, then the correspondent perturbation theory
\cite{CMIM,GKSpri} modifies the model for open system.

The inertial manifold \cite{IneManCFTe88,IneManTe88,Kev} is the
manifold where the limit behaviour of the system is located; it
exponentially attracts motions when $t\rightarrow \infty$. For a
closed system the equilibrium (one point) is already the inertial
manifold. In the theory of inertial manifolds the estimates of
inertial manifolds dimension for several classes of (open) systems
are created and finiteness of this dimension is proved. Inertial
manifolds could be considered as the lowest level of any hierarchy
of slow manifolds. They belong to all the slow invariant manifolds
of the hierarchy. In our construction we build the hierarchy of
infinite--dimensional approximate slow manifolds for the Boltzmann
equation and do not try to find the smallest invariant manifolds
for open systems.

And, finally, we should ask the question: what chain is better,
could we prove that the second hierarchy with scattering rates
instead of usual moments is better than the standard Grad
hierarchy? We cannot prove this exactly, but can only argue
plausibly that the second hierarchy should lead to dynamic
invariance faster, than the first one, and support this point of
view by examples.

\end{document}